\documentclass[12pt,preprint]{aastex} 

\slugcomment{Accepted for publication in the Astrophysical Journal.}
\shorttitle{The Nature of the SS 433 Binary}
\shortauthors{Marshall et al.}

\begin{document}
\title{Multiwavelength Observations of the SS 433 Jets}
\author{Herman L. Marshall\altaffilmark{1},
Claude R. Canizares\altaffilmark{1},
Todd Hillwig\altaffilmark{2},
Amy Mioduszewski\altaffilmark{3},
Michael Rupen\altaffilmark{3},
Norbert S. Schulz\altaffilmark{1},
Michael Nowak\altaffilmark{1},
Sebastian Heinz\altaffilmark{4}}
\altaffiltext{1}{Kavli Institute for Astrophysics and Space Research,
 Massachusetts Institute of Technology, 77 Massachusetts Ave.,
 Cambridge, MA 02139, USA}
\altaffiltext{2}{Dept. of Physics and Astronomy, Valparaiso University, IN, 46383}
\altaffiltext{3}{NRAO, PO Box 2, Socorro, NM, 87801}
\altaffiltext{4}{Astronomy Dept., 5408 Sterling Hall, University of Wisconsin, Madison, WI, 53706}
\email{hermanm@space.mit.edu,
crc@space.mit.edu,
todd.hillwig@valpo.edu,
amiodusz@nrao.edu,
mrupen@aoc.nrao.edu,
nss@space.mit.edu,
mnowak@space.mit.edu,
heinzs@astro.wisc.edu}

\begin{abstract}
We present observations of the SS 433 jets using
the {\em Chandra} High Energy Transmission Grating Spectrometer
with contemporaneous optical and VLBA observations.
The X-ray and optical emission line regions are found to be related but not
coincident as the optical line emission persists for days while the X-ray
emission lines fade in less than 5000 s.  The line Doppler shifts from the
optical and X-ray lines match well, indicating that they are less
than $3 \times 10^{14}$ cm apart.
The jet Doppler
shifts show aperiodic variations that could result from shocks in interactions
with the local environment.
These perturbations are consistent with
a change in jet direction but not jet speed.
The proper motions of the radio knots match the kinematic model
only if the distance to SS 433 is 4.5 $\pm$ 0.2 kpc.
Observations during eclipse
show that the occulted emission is very hard, seen only above 2 keV and
rising to comprise $>50$\% of the flux at 8 keV.
The soft X-ray emission lines from the jet are not blocked,
constraining the jet length to $\ga 2 \times 10^{12}$ cm.
The base jet density is in the range
$10^{10-13}$ cm$^{-3}$, in contrast to our previous estimate based
on the Si {\sc xiii} triplet, which is
likely to have been affected by UV de-excitation.
There is a clear overabundance of Ni by a factor of about 15
relative to the solar value, which may have resulted
from an unusual supernova that formed the compact object.
\end{abstract}


\section{Introduction}

SS 433 is still the only source that is known to have emission lines
from ionized gas in a highly collimated jet from a compact object.
Thus, it is an important
member of the ``microquasar'' class of X-ray binaries
\cite[cf.][]{1999ARA&A..37..409M}.
For an early review of the
source properties, see \cite{1984ARA&A..22..507M} or a more recent
review by  \cite{2004ASPRv..12....1F}.  The so-called ``kinematic model''
describes the Doppler shifts of the blue- and redshifted H$\alpha$
lines as a pair of oppositely directed, precessing jets
with speed, v$_j = 0.26c$ \citep{1989ApJ...347..448M}.
The model has been used for over 30 years when analyzing
SS 433 data.  There have been some updates to the parameters
and to the model,
most notably to add nutation with a period which is about half of the orbital
period and to add small jet velocity variations
at the orbital period \citep{2005ApJ...622L.129B}.
Based on radio images, \citet{2005ApJ...622L.129B} also determined that
the distance is 5.5 $\pm$ 0.2 kpc for a scale of
0.027 pc/\arcsec\ but \citet{2002MNRAS.337..657S} argue that the distance
should be somewhat smaller: 4.61 $\pm$ 0.35 kpc.

The precise
description of the line positions was given by \citet{1989ApJ...347..448M}: the jet
orientation precesses with a
162.5 day period in a cone with half-angle 19.85\arcdeg\ about
about an axis which is 78.83\arcdeg\ to the line of sight.
The lines from the jets are Doppler shifted
with this period so that the maximum
redshift is about 0.15 and the most extreme blueshift is about -0.08.
We will use the reference phases and periods from \citet{1998vsr..conf..103G},
where the precession reference phase and precession period are JD  2451458.12
\cite[updated by][]{2002ApJ...566.1069G}
and 162.15 d; eclipses occur at the 13.08211 d binary period
with reference phase of JD 2450023.62; and the jets nutate
with period of 6.2877 d, ephemeris of JD 2450000.94,
and an amplitude $z_{\rm nut} = 0.009$ \citep{2002ApJ...566.1069G}.
For the remainder of this paper, we will use truncated Julian date, defined
by TJD $\equiv$ JD $-$ 2450000.
Assuming uniform outflow,
the widths of the optical lines were used to estimate the
opening angle of the jet: $\sim$5\arcdeg\ \citep{1980ApJ...238..722B}.
For more details of the optical spectroscopy, see the reviews by
\citet{1984ARA&A..22..507M} and \citet{2004ASPRv..12....1F} .
 
SS 433 has radio jets oriented at a position angle of
100\arcdeg\ (east of north) which show an oscillatory pattern
that can be explained by helical motion of material flowing
along ballistic trajectories \citep{1981ApJ...246L.141H}.
Using VLBI observations of ejected knots, \citet{1993A&A...270..204V}
confirmed the velocity of the jet and
the kinematic model, although deviations of up to 10\%
were found from a sequence of observations spanning
40 days \citep{2003AAS...203.3105M,2004AAS...20510401S}.
The radio jets extend from the milliarcsec scale to several
arc seconds, a physical range of $10^{15-17}$ cm from the
core.
The optical emission lines originate
in a region $\la 3 \times 10^{15}$ cm across, based
on light travel time arguments \citep{1980ApJ...241.1082D}.

There is substantial literature on the X-ray emission from SS 433.
\citet{1979ApJ...230L.145M} were the first to demonstrate that SS 433 is an X-ray
source.
\citet{1986MNRAS.222..261W} showed that the X-ray lines were
Doppler shifted, locating at least some of the X-ray emission in the
jets.
Several groups used eclipse observations to estimate or limit the
length of the X-ray emitting portion of the jet, obtaining values
of $\sim 10^{12}$ cm
\citep{1987MNRAS.228..293S,1989PASJ...41..491K,1991A&A...241..112B,1992AZh....69..282A}.
More recently, the jet lengths were modeled to obtain the binary
mass ratio $q \equiv M_X/M_C \sim 0.3$, where $M_X$ is the mass
of the compact object and $M_C$ is the mass of the companion
star \citep{2009MNRAS.397..479C}.
\citet{1991A&A...241..112B}
and \citet{1996PASJ...48..619K} developed the model that
the X-ray emission originates in thermal gas near the bases of the jets.
The outflowing, cooling gas
adiabatically expands and cools until $kT$ drops to
about 100 eV at which point the jet is thermally unstable
\citep{1988A&A...196..313B}.
\cite{1996PASJ...48..619K} found that the redshifted Fe {\sc xxv} line
was fainter than the blueshifted line, after correcting for the expected
Doppler diminution, leading them to
conclude that the redward jet must be obscured by neutral material
in an accretion disk.

\begin{figure}
\includegraphics[width=15cm]{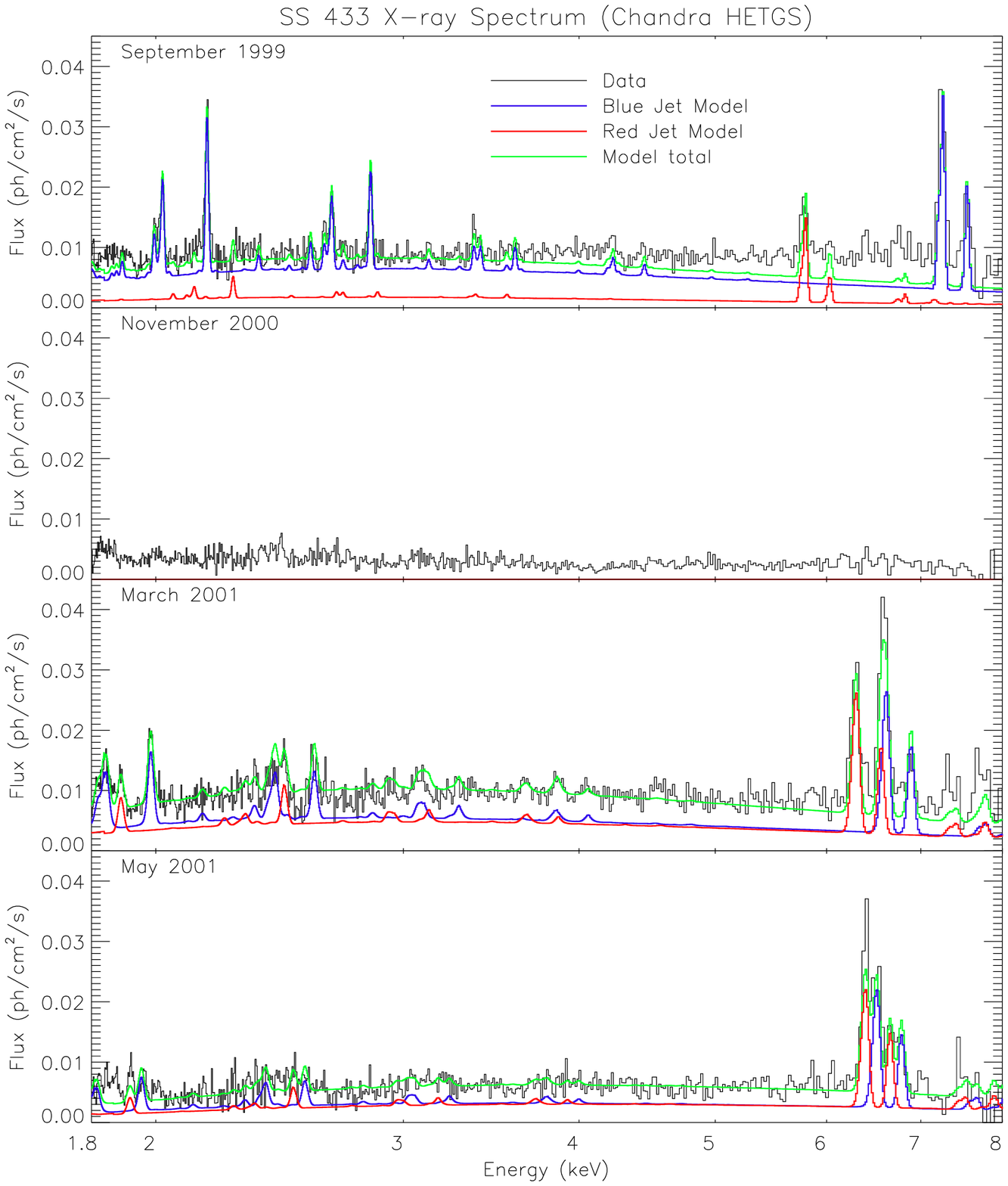}
\caption{Four X-ray spectra of SS 433 taken with the {\em Chandra} High
Energy Transmission Grating Spectrometer. The models were taken
from \cite{2002ApJ...564..941M} and \cite{2006lopez} for the
1999 and March 2001 data sets.  The model for the May 2001
data set was based on the model for March 2001 except that
the Doppler shifts were changed to 0.0460 and 0.0265
\citep{2003PASJ...55..281N} and a factor of $0.85(E/1 {\rm keV})^{-0.5}$ was
applied to reduce the continuum and the flux in the soft band.  There
are no clear emission lines in the November 2000 spectrum.
  }
\label{fig:4spectra}
\end{figure}

\citet{2002ApJ...564..941M} (hereafter, Paper I) used an observation of SS 433
with the {\em Chandra} High Energy Transmission Grating Spectrometer (HETGS)
to resolve the X-ray lines, detect fainter lines than previously observed,
and measure lower energy lines that could not be readily detected
in the {\em ASCA} observations.
In Paper I, blueshifted lines dominated
the spectrum and the redshifted lines were all relatively weak
by comparison, so most work focused on modelling the blue jet
(see Fig.~\ref{fig:4spectra}).
The jet Doppler shifts were determined very accurately due to
the small scatter of the individual lines and the narrowness of their
profiles, indicating that all line-emitting gas flows at the same speed.
The jet bulk velocity was $\beta$c, where $\beta$=$0.2699\pm0.0007$.
This jet velocity was larger than the velocity inferred from optical
emission lines by $2920\pm440$ km s$^{-1}$.
Gaussian fits to the emission lines were all consistent with the
same Doppler broadening: $1700\pm80$ km s$^{-1}$ (FWHM).
Relating this broadening to the maximum velocity due to beam
divergence, the opening angle of the jet is
$1.48^{\circ}\pm0.07^{\circ}$.\footnote{Due to a coding error
in Paper I, the opening angle was incorrectly reported to
be $1.23^{\circ}\pm0.06^{\circ}$.  We thank Dr.\ Rosario
Iaria for pointing out this error.}

\citet{2006lopez} presented two more HETGS observations
of SS 433; one showed no lines (Nov. 2000)
and the other was taken during eclipse (see Fig.~\ref{fig:4spectra}).
In the eclipse observation (from March 2001), lines from both the red- and
blueshifted jets were clearly detected and were of
comparable strength.
Previous X-ray observations of the SS 433 system during
eclipse did not provide good
detections of emission lines from the receding jet,
leading to the suggestion that the entire inner region
of the jets was eclipsed. However, the HETGS
observation shows numerous emission lines from both jets.

\citet{2003PASJ...55..281N} published the HETGS spectra from
observation on 2001 May 12 (see Table~\ref{tab:observations}), the third of
a set of three observations.
Based on the width of the most highly ionized and blended Fe lines, they concluded
that the base of jet subtends a larger angle than the remainder.
Fig.~\ref{fig:4spectra} shows that the combined spectrum from the
data set shows highly blended lines and very weak low energy lines.
The authors did not say why only the third observation was analyzed
but it seems likely that it was because the source was somewhat
fainter during observations 1940 and 1941 (See
Table~\ref{tab:observations} and Fig.~\ref{fig:lc}).

Here, we present new observations of SS 433 using the HETGS
as well as optical, hard X-ray, and VLBA observations
obtained contemporaneously.
The observations were designed to provide a high signal
spectrum for testing the emission model; to obtain spectra during eclipse
to test the obscuration of the red jet's emission; to accumulate sufficient
observing time to detect nutational and precessional changes
of the jets' Doppler shifts; and to obtain comparison data in other wavebands.
Some preliminary results were reported
earlier \citep{2006hrxs.confE..20M,2008ralc.conf..454M}, calling attention to
a Doppler shift change over a 20 ks interval, much shorter than
the nutational, orbital, or precession periods.

\section{Observations and Data Reduction}

\subsection{{\em Chandra} X-ray Observations}

\begin{figure}
\includegraphics[width=15cm]{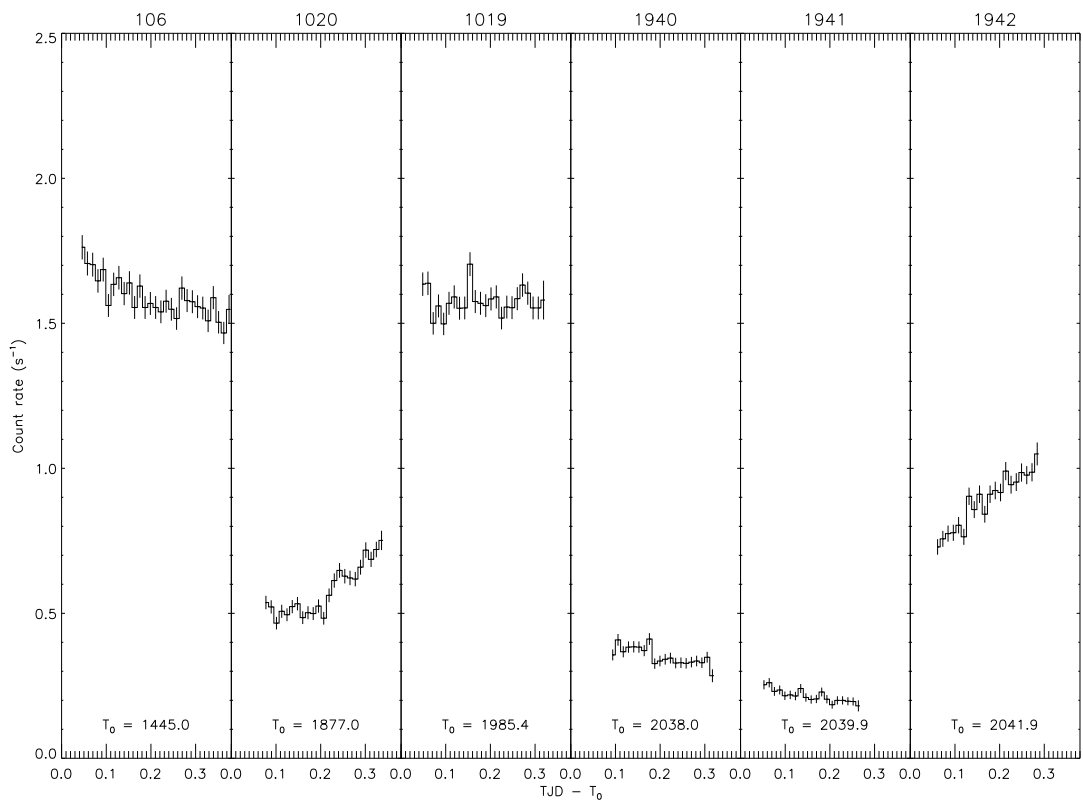}
\caption{Light curves of SS 433 from the first six {\em Chandra} HETGS
observations (see Table~\ref{tab:observations}),
using the MEG count rates (0.5-8 keV) in 1000 s bins
from TGCat \citep{1538-3881-141-4-129}.
The observation ID is given at the top of each panel.
The X-ray source was in eclipse during observations 1019 and 1940,
showing that the MEG count rates vary intrinsically by almost a factor of 10
and that the rates are not always lower during eclipses.
A brightening by a factor of 4 occurred in less than two days
between observations 1941 and 1942.
  }
\label{fig:lc}
\end{figure}

\begin{figure}
\includegraphics[width=15cm]{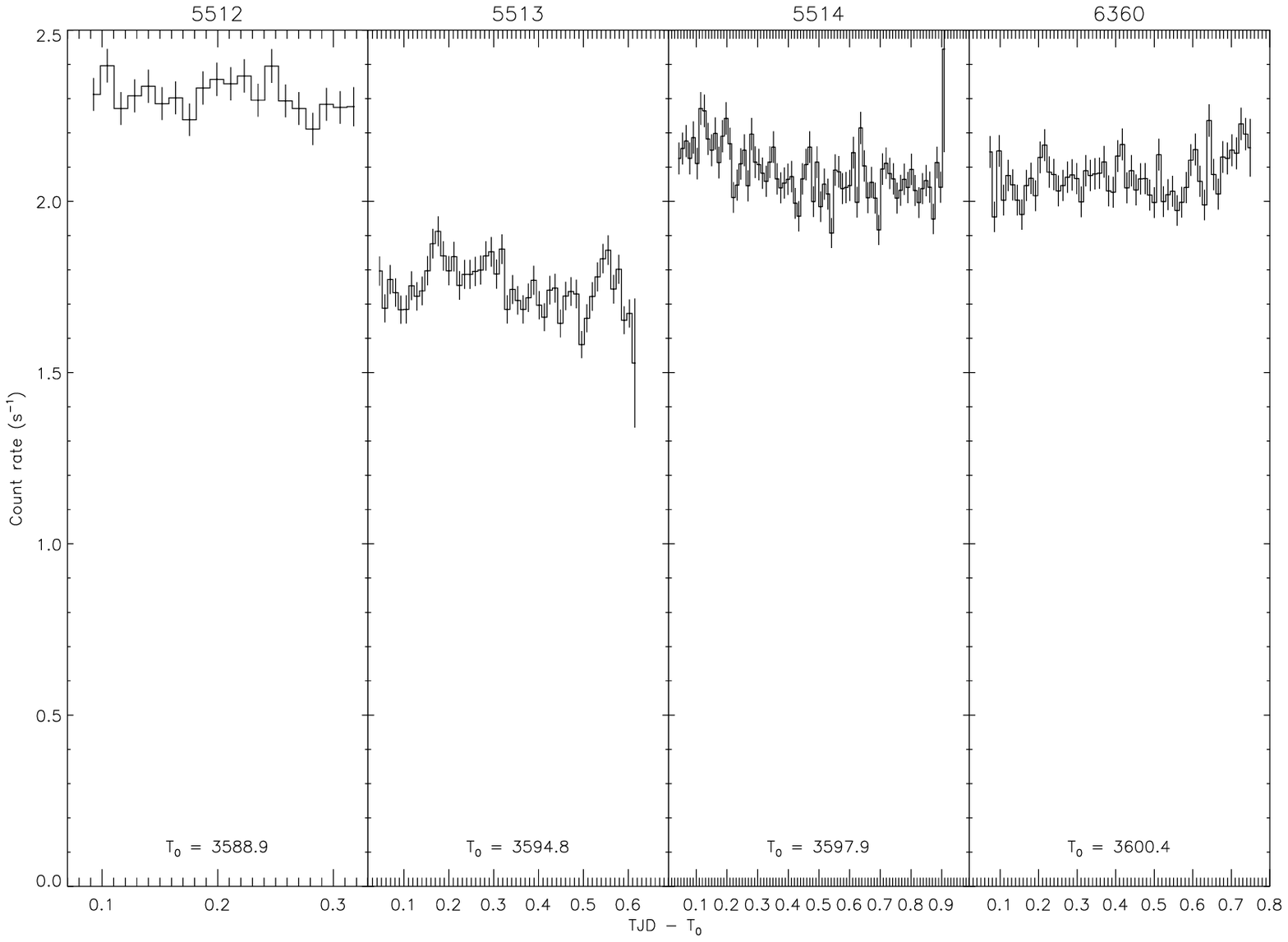}
\caption{Same as Fig.~\ref{fig:lc} but for the
August 2005 {\em Chandra} observations.
The source was in eclipse during observation 5513 but was
otherwise constant within $\pm$ 7\%.
  }
\label{fig:lc2005}
\end{figure}

SS 433 was observed with the HETGS several times in
August 2005, as given in Table~\ref{tab:observations}.
Included in Table~\ref{tab:observations} are all the previous
HETGS observations, with the count rate in the medium energy
grating (MEG) band as well as the precession and orbital
phases.
The grating data and the count rates were extracted from
the Transmission Grating Catalog \citep[TGCat,][]{1538-3881-141-4-129}
as well as the light curves in 1000 s bins, shown in Fig.~\ref{fig:lc}
(for the first 6 observations) and Fig.~\ref{fig:lc2005} for
the 2005 observations.
The count rate varies by almost a factor of 10 from year to year
and by a factor of 2-4 on a time scale of days.  The lowest count
rate is not found during eclipse.  During the 2005 observations,
however, the total count rate varies by less than $\pm$7\% except
during the eclipse, where the rate is about 15\% low.

\begin{deluxetable}{lrcccccl}
\tablecolumns{5}
\tablewidth{0pc}
\tabletypesize{\scriptsize}
\tablecaption{{\em Chandra} Spectroscopic Observations of SS 433 \label{tab:observations} }
\tablehead{
	\colhead{Date} & \colhead{Start TJD\tablenotemark{a}} & \colhead{Observation ID} &
	\colhead{Exposure} & \colhead{MEG Rate} & \colhead{$\phi_{\rm prec}$} &
	\colhead{$\phi_{\rm orb}$} & \colhead{Ref} \\
	\colhead{} & \colhead{} & \colhead{} & \colhead{(ks)} & \colhead{(cps)} & \colhead{} & \colhead{} & \colhead{}
	 }
\startdata
1999 Sep 23 & 1445.04 & ~106 & 28.7 & 1.59 & 0.92 & 0.67 & \cite{2002ApJ...564..941M} \\
2000 Nov 28 & 1877.07 & 1020 & 22.7 & 0.58 & 0.58 & 0.69 & \cite{2006lopez} \\
2001 Mar 16 & 1985.44 & 1019 & 23.6 & 1.58 & 0.25 & 0.97 & \cite{2006lopez}  \\
2001 May 8 & 2038.09 & 1940 & 19.6 & 0.35 & 0.58 & 0.00 & this work \\
2001 May 10 & 2039.95 & 1941 & 18.5 & 0.22 & 0.59 & 0.14 & this work \\
2001 May 12 & 2041.96 & 1942 & 19.7 & 0.88 & 0.60 & 0.29 & \cite{2003PASJ...55..281N} \\
2005 Aug 6 & 3588.97 & 5512 & 21.0 & 2.31 & 0.14 & 0.54 & this work \\
2005 Aug 12 & 3594.84 & 5513 & 48.2 & 1.75 & 0.18 & 0.01 & this work \\
2005 Aug 15 & 3597.94 & 5514 & 73.1 & 2.08 & 0.20 & 0.25 & this work \\
2005 Aug 17 & 3600.45 & 6360 & 59.1 & 2.07 & 0.21 & 0.44 & this work \\
\enddata
\tablenotetext{a}{TJD is defined as JD $-$ 2450000.}
\end{deluxetable}

Due to the lack of lines in observation ID 1020, a short ``trigger''
observation was executed on 2005 Aug 6 (obsID 5512).
A quick look analysis showed significant emission lines, so the
remaining observations were scheduled to start about a week later.
Observation 5513 was taken during eclipse when the companion blocked
part of the X-ray continuum.
The predicted angle of the jet to the line of sight, $\alpha$, varied
from 69\arcdeg\ to 77\arcdeg\ during the 2005 observations.
Within each observation, the data are essentially continuous.
Light curves from all observations are included in Fig.~\ref{fig:lc2005}.

\subsection{{\em Rossi} X-ray Timing Explorer (RXTE) Observations}

\label{sec:xte}

\begin{figure}
\includegraphics[angle=90,width=15cm]{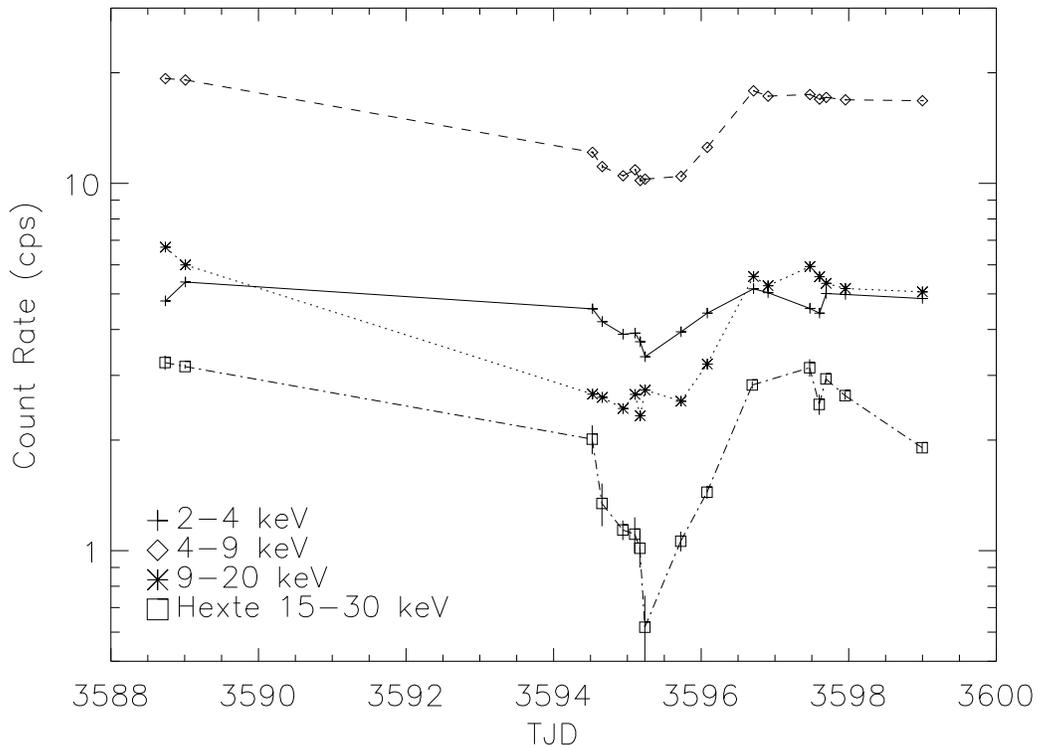}
\caption{Count rates from {\em RXTE} observations taken in conjunction
 with the {\em Chandra} HETGS.
 Three broad PCA bands and one HEXTE band are plotted for each observation ID
 listed in Table~\ref{tab:xte}.
 Error bars are purely statistical and do not reflect variations within data sets.
 Eclipse center is at TJD 3595.036, near the time when each band is low.
 Eclipse depth increases with energy, from 19\% in the 2-4 keV band
 up to 57\% in the 15-30 keV band (see text).
  }
\label{fig:xte}
\end{figure}

RXTE observations were obtained to overlap
the 2005 {\em Chandra} HETGS observations.
The observation time periods are given in Table~\ref{tab:xte}.
For each observation ID, the average count rates were determined
from the standard data products in several bandpasses.
These are shown in Fig.~\ref{fig:xte}.
The count rates in each band are within 5\% of being
constant outside of eclipse, which is centered at TJD 3595.036 and
spans about two days.  Taking the eclipse to be from
TJD 3594 to 3596.5, we can determine
the average flux diminution relative to rates out of eclipse, from
TJD 3596.5 to 3598.5.
We find that the count rates dropped by $19.2 \pm 0.4$\%,
$35.9 \pm 0.2$\%, $49.8 \pm 0.5$\%, and $57 \pm 3$\% for
the 2-4, 4-9, 9-20, and 15-30 keV bands, respectively.
For the HEXTE 30-60 keV band, we obtained average
count rates of $0.29 \pm 0.04$ cps in eclipse and
$0.958 \pm 0.044$ outside of eclipse, for a 70\%
decrease during eclipse.

\begin{deluxetable}{ccr}
\tablecolumns{3}
\tablewidth{0pc}
\tabletypesize{\scriptsize}
\tablecaption{{\em Rossi} XTE Observations of SS 433 in 2005\label{tab:xte} }
\tablehead{
	\colhead{Observation ID} & \colhead{Start Date} & \colhead{Exposure} \\
	\colhead{} & \colhead{(TJD)} & \colhead{(ks)}
	 }
\startdata
91092-01-01-00 & 3588.72 &  1.69 \\
91092-01-02-00 & 3588.85 & 26.10 \\
91092-02-01-02 & 3594.52 &  1.30 \\
91092-02-01-04 & 3594.65 &  1.90 \\
91092-02-01-00 & 3594.89 & 10.27 \\
91092-02-01-01 & 3595.09 &  1.71 \\
91092-02-07-01 & 3595.16 &  1.79 \\
91092-02-01-03 & 3595.22 &  2.01 \\
91092-02-02-00 & 3595.67 &  9.84 \\
91092-02-03-00 & 3596.00 & 17.82 \\
91092-02-04-00 & 3596.61 & 13.74 \\
91092-02-05-00 & 3596.85 & 11.41 \\
91092-02-06-01 & 3597.46 &  1.51 \\
91092-02-06-02 & 3597.59 &  1.17 \\
91092-02-06-00 & 3597.64 &  9.53 \\
91092-02-07-00 & 3597.90 & 10.67 \\
91092-02-08-00 & 3598.88 & 22.61 \\
\enddata
\end{deluxetable}

\subsection{Optical Observations}

\begin{figure}
\includegraphics[width=15cm]{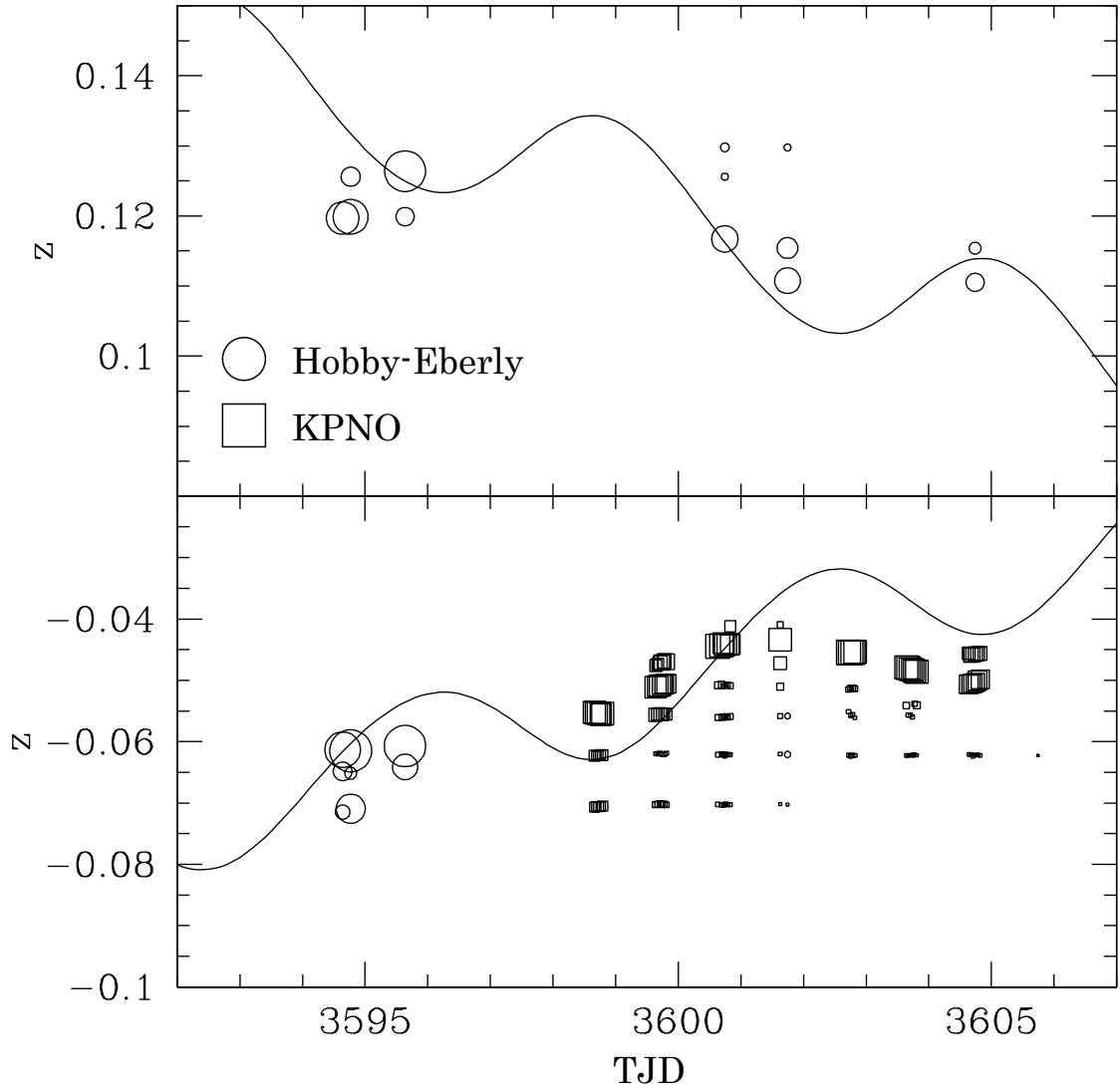}
\caption{
 Doppler shifts from optical observations taken during the 2005 campaign
 as a function of TJD.  Several spectra were obtained per night
 at KPNO.  For most observations, the lines were fitted with several Gaussians
 at different Doppler shifts.
 The size of the symbol relates to the line strength.
 The lines provide a nominal ephemeris from the kinematic model
 of the jet Doppler shifts, with a time delay of 4 days.
  }
\label{fig:optical}
\end{figure}

Simultaneous optical spectra were obtained in order to model
the relationship between the various emission regions.
Data were obtained
at the Hobby-Eberly Telescope (HET) with the Medium Resolution
Spectrograph with 2\arcsec\ fibers (with spectral resolution $R \approx 5000$) and
at the KPNO 2.1 m telescope using
GoldCam with a 2\arcsec\ slit and grating 56 ($R \approx 4000$).
 The HET spectra were obtained generally only once per night while
there were up to six KPNO spectra each night, at 40-50 min intervals.
Doppler shifts of the H-alpha lines are shown in Fig.~\ref{fig:optical}.
A model of the Doppler shift variation is computed from the standard kinematic
model but with a precessional time delay of 4 d, estimated
to provide a better fit to the dominant lines.  The effect of the time delay is to shift
the precession model's expected Doppler shifts later in time.
Precession phase jitter of
5-10 d -- either later or earlier -- has been found in many
optical observations \citep{2001ApJ...561.1027E}.

As found by
\citet{1993A&A...270..204V}, emission lines appear at Doppler shifts that initially
match the kinematic model (after accounting for phase jitter)
but then persist for many days.
Doppler shift
deviations from the kinematic model can be of order 3000 km s$^{-1}$.
Optical line FWHM values were 10-50 \AA, giving Gaussian velocity
dispersions of 200-1000 km/s.  Emission line positions and
equivalent widths were
calculated using up to seven Gaussians with different
centroids for lines that were clearly resolved.  Line equivalent
widths varied from 0.17 \AA\ to 52 \AA.
Strongly blended lines were often
fit using a single Gaussian.  In these cases the line centers were not
appreciably different than found by fitting several unresolved Gaussians.
Likewise, most blended lines are dominated by a single component and
thus the equivalent widths are also not greatly affected by a single fit.

\subsection{VLBA Observations}

Very long baseline array (VLBA) observations at 15.4 GHz were obtained
daily during part of the campaign from TJD 3591.7 to 3606.7, in three hour periods.
The data were taken by alternating 2 min observations of SS 433 with
1 min on the phase calibrator J1912$+$0518.
Because the phase calibrator was only 0.42\arcdeg\ away from SS 433, absolute
astrometry should be good to better than 0.2 mas.
Calibrators J1800$+$7828 and J1907$+$0127 were also observed occasionally
as check sources.
The frequency range from 15.3495 GHz to 15.3810 GHz was covered in 8 MHz bandpasses. 
The data were reduced in the standard manner using NRAO's data
reduction package, AIPS \citep{2003ASSL..285..109G}.
Images were made with natural weighting, without the necessity of self-calibration.

\begin{figure}
\includegraphics[width=14cm]{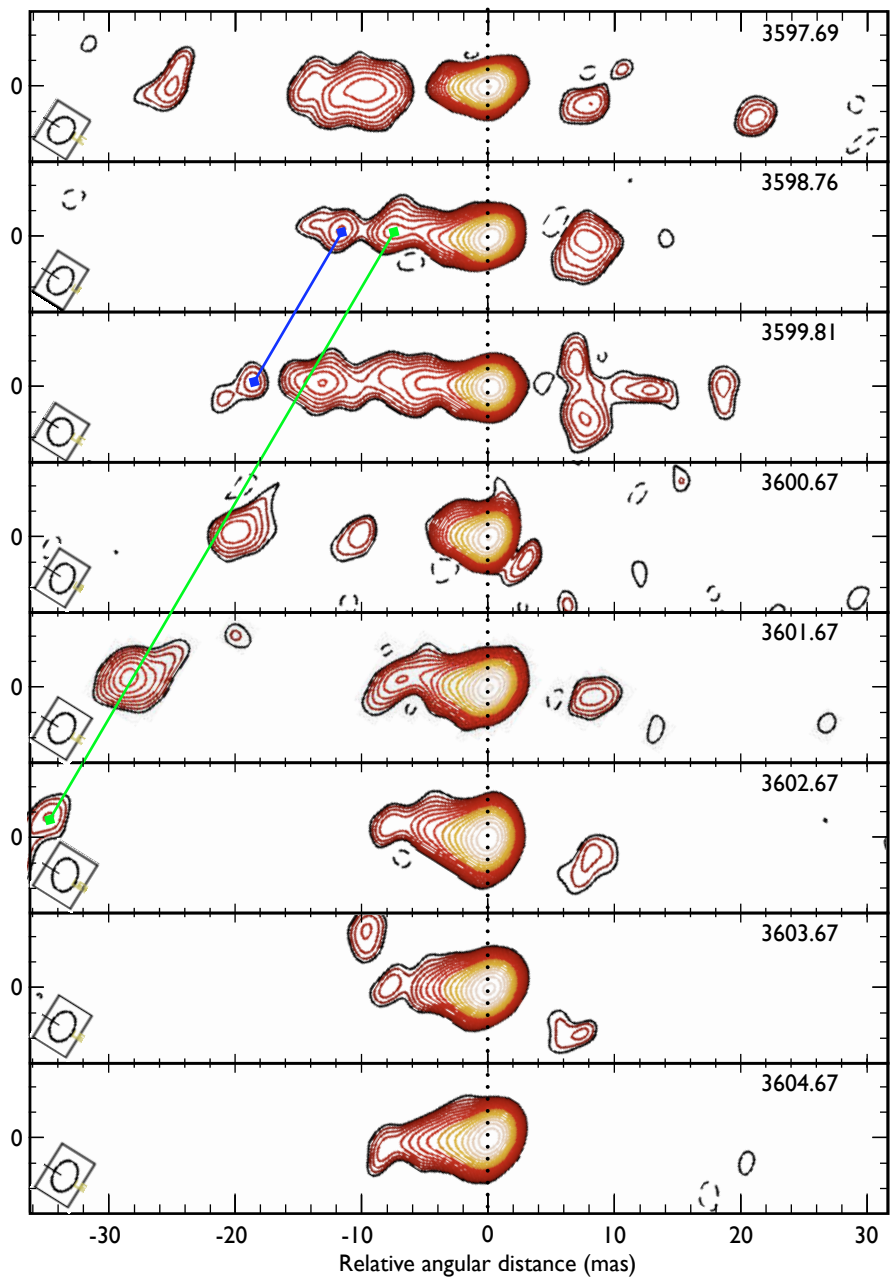}
\caption{
 Images from the VLBA campaign rotated -32$\arcdeg$ so that the
 jet is approximately horizontal.  The
 beam is shown in the lower left corner of each image; the short line
 indicates E for each beam.
 Observation dates are given in TJD for each panel.
 Angular distances are marked in milli-arcseconds
 relative to the map peak, whose astrometric position is
 consistent from image to image to within 0.2 mas.
 The lowest contour is 0.5 mJy/beam, with one negative
 contour at -0.5 mJy/beam (dashed), and contours increase
 by factors of $2^{1/2}$; maxima are between 60 and 100 mJy/beam.
 Green and blue lines connect components that
 were measured and shown in Fig.~\ref{fig:vlbacomponents}.
  }
\label{fig:vlbaimage}
\end{figure}

\begin{figure}
\includegraphics[angle=90,width=15cm]{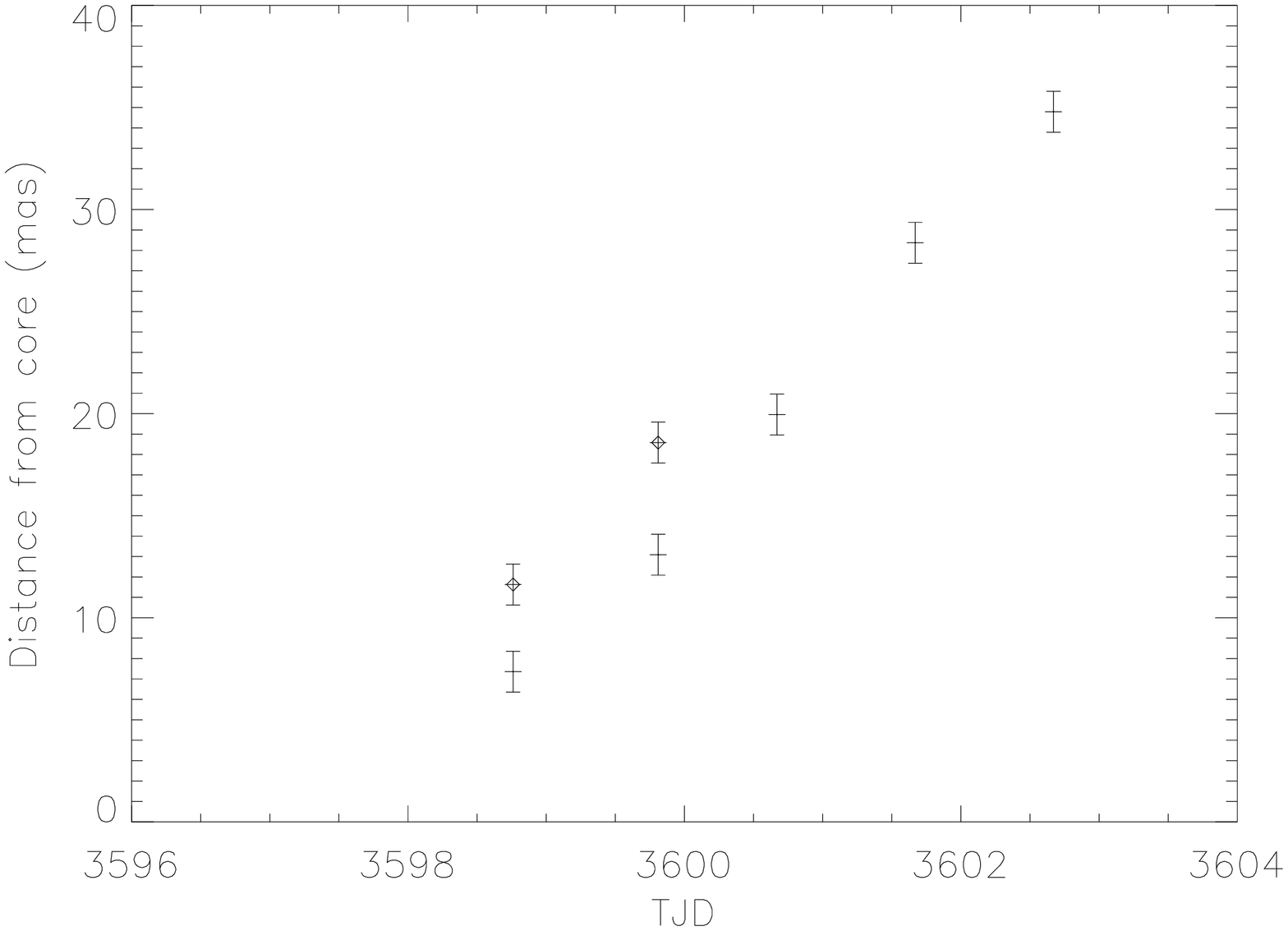}
\caption{
 Distances (in milli-arcseconds) from the core for
 two components in the VLBA images shown in Fig.~\ref{fig:vlbaimage}.
 Linear trends were fit to obtain the knots' proper motions and ejection
 dates.  Horizontal lines span the observation periods, which were 3hr
 for each observation.
  }
\label{fig:vlbacomponents}
\end{figure}

Fig.~\ref{fig:vlbaimage} shows 8 images from the campaign.
The beam is about 2 mas in the N-S direction
and 1 mas in the E-W direction.
The core position was stable to within astrometric accuracy
at $(\alpha, \delta) = (19^h11^m49^s.56824, +04\arcdeg58\arcmin57\farcs7632)$ (epoch J2000).
The position angle of the jet is about 122$\arcdeg$ east of north,
so the images were rotated -32$\arcdeg$ to appear horizontal
in the figure and referenced to the core.
Two knots were tracked in the images; measurements are shown
in Fig.~\ref{fig:vlbacomponents}.
The measured proper motions were $5.9 \pm 0.6$ mas/day,
launched at TJD 3596.6 $\pm$ 0.1, and
$7.14 \pm 0.32$ mas/day, launched at TJD 3597.72 $\pm$ 0.05.
Given uncertainty in locating the core, we include a possible
systematic error on the launch dates of $\pm 0.25$ d.

\subsection{X-ray Doppler Shifts}
\label{sec:dopplershifts}

\begin{figure}
\includegraphics[angle=270,width=16.5cm]{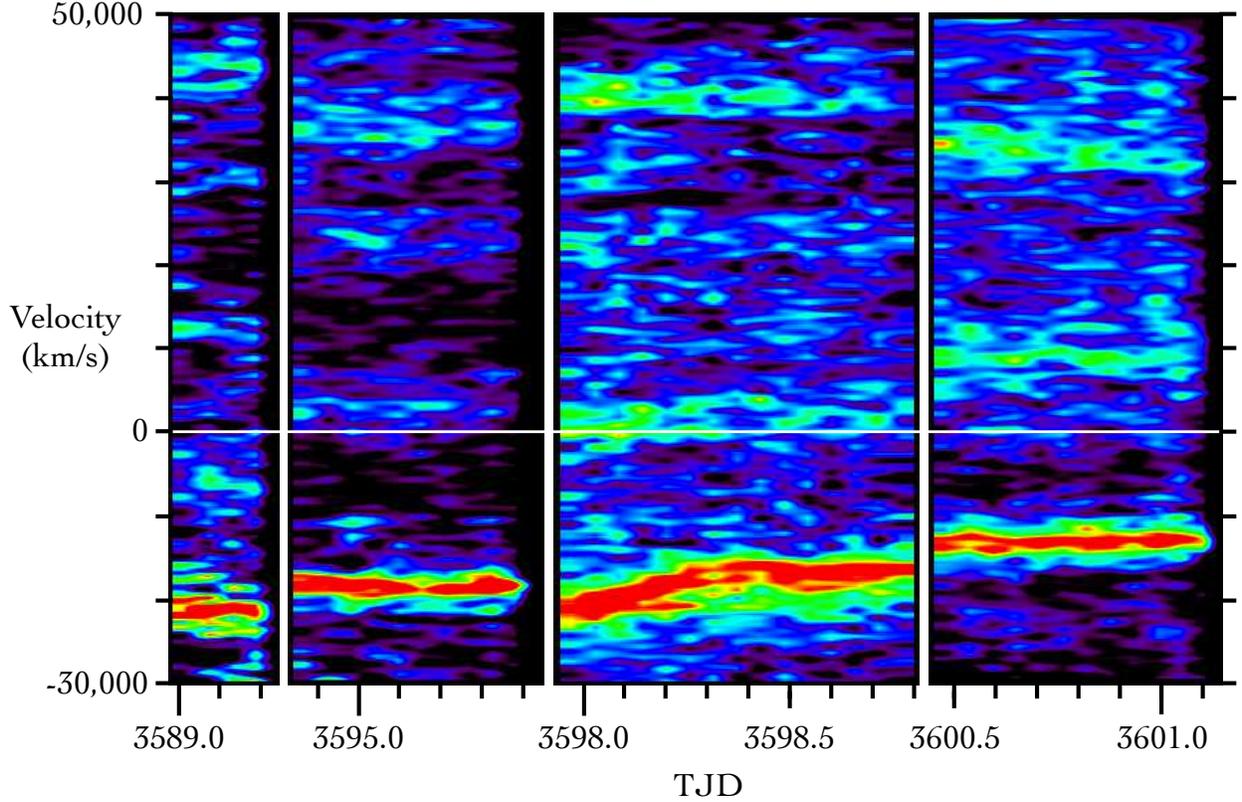}
\caption{Trailed spectra from the {\em Chandra} observations.
 The lines of Mg {\sc xii}, Si {\sc xiv}, Fe {\sc xxvi}, and Fe {\sc xxv}
 were combined in 500 km/s velocity bins at 5 ks intervals and
 slightly smoothed for visualization.  The red track follows the Doppler
 shift of the blueshifted emission lines near -20,000 km/s; the redshifted lines
 are barely visible between 30,000 km/s and 50,000 km/s.
 The weak features in the 0-10,000 km/s observed after TJD 3597 are
 due to weaker, adjacent lines in the spectra.
 Note that the blueshift scarcely varies during any given observation except
 near TJD 3598.0, where there is a clear linear
 increase of the blueshift by about 5000 km/s in a span of 0.3 d.
 The emission lines are not appreciably broadened
 temporally during the linear blueshift increase, indicating that most of the gas
 cools in less than 5000 s.
 The redshift barely changes during the observations
 but all Doppler shifts changed by about 3000 km/s between observations.
  }
\label{fig:trailed}
\end{figure}

From Fig.~\ref{fig:optical}, we expect that the Doppler shifts measured
from the X-ray spectra will vary significantly over the several days covered
by the {\em Chandra} observations.
For this reason, the X-ray data were divided into 5 ks intervals
to generate a series of spectra.
Based on the results in Paper I, we expect that all the strong emission
lines have the same Doppler shifts, so that the data for different lines
were combined to improve the signal/noise.
For specific, strong emission lines with rest energy $E_0$, the X-ray events
from an energy range $[0.9 E_0,1.17 E_0]$ were accumulated in 500 km/s bins to
produce a velocity profile every 5 ks.  The Ly$\alpha$ lines of H-like Mg {\sc xii},
Si {\sc xiv}, and Fe {\sc xxvi} were chosen, along with the K$\alpha$ line
from He-like Fe {\sc xxv}.
The result, in Fig.~\ref{fig:trailed}, shows that 1) the blueshifted lines are substantially
stronger than the redshifted lines, 2) the Doppler shifts mostly remain constant
during 20-50 ks observations, 3) the Doppler shifts change by $\ga$ 3000 km/s
during the day or more between observations, and, surprisingly, 4) a 5000 km/s
shift was observed during a 0.3 d period starting at about TJD 3598.0. 

\begin{figure}
\includegraphics[angle=90,width=17cm]{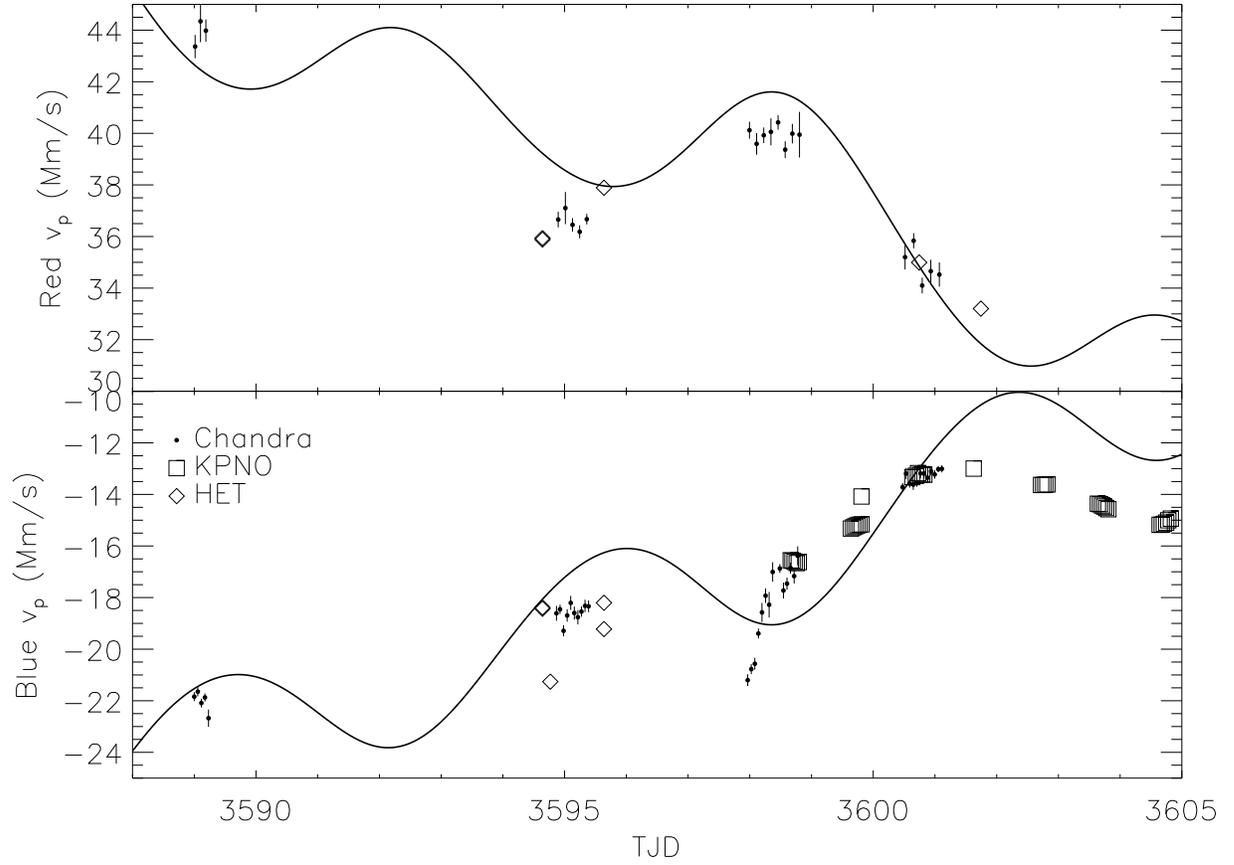}
\caption{
 Doppler shifts from optical observations compared to the
 values obtained from {\em Chandra}.
 The Doppler shift model is the nominal ephemeris from the
 kinematic model, with a time delay of 4 days as plotted in Fig.~\ref{fig:optical}.
 Only the Doppler shifts for the strongest optical lines are shown, for
 clarity of comparison to the X-ray measurements.
 The optical and X-ray Doppler shifts generally agree within
 the uncertainties, when the results overlap.
  }
\label{fig:optxray}
\end{figure}

Comparing the optical and X-ray Doppler shifts
(Fig.~\ref{fig:optxray}), we find that the Doppler shifts match to
$<$ 1000 km s$^{-1}$.
We can set an upper limit of 0.40 d for the maximum time discrepancy between
the optical and X-ray measurements of the blueshifted jet by noting that the
X-ray measurements before TJD 3598.26 were all significantly more negative
than the next optical measurement at TJD 3598.66.  For the redshifted
jet, the limit is obtained from the last optical measurement, at
TJD 3601.75, which is significantly different from the last X-ray measurement,
at TJD 3601.07, for an upper limit to the delay of 0.68 d.

As in Paper I, we assumed the system is comprised of two perfectly opposed
jets at an angle $\alpha$ to the line of sight.
Then, the Doppler shifts of the blue and red jets are given by 

\begin{equation}
\label{eq:redshift}
 z = \gamma (1 \pm \beta \mu ) - 1
\end{equation}

\noindent
where v$_j = \beta c$ is the velocity of the jet flow,
$\gamma = (1 - \beta^2)^{-1/2}$, and $\mu = \cos \alpha$.
As in the first paper, the line Doppler shifts were used to
obtain an estimate of the $\gamma$ by adding
the redshifts to cancel the $\beta \mu$ terms
and $\mu$ by subtracting redshifts:

\begin{eqnarray}
\label{eq:gamma}
\gamma = \frac{z_b + z_r}{2} + 1 \\
\mu = \frac{z_r - z_b}{2 \gamma \beta}
\label{eq:alpha}
\end{eqnarray}

\begin{figure}
\center
\includegraphics[width=13cm]{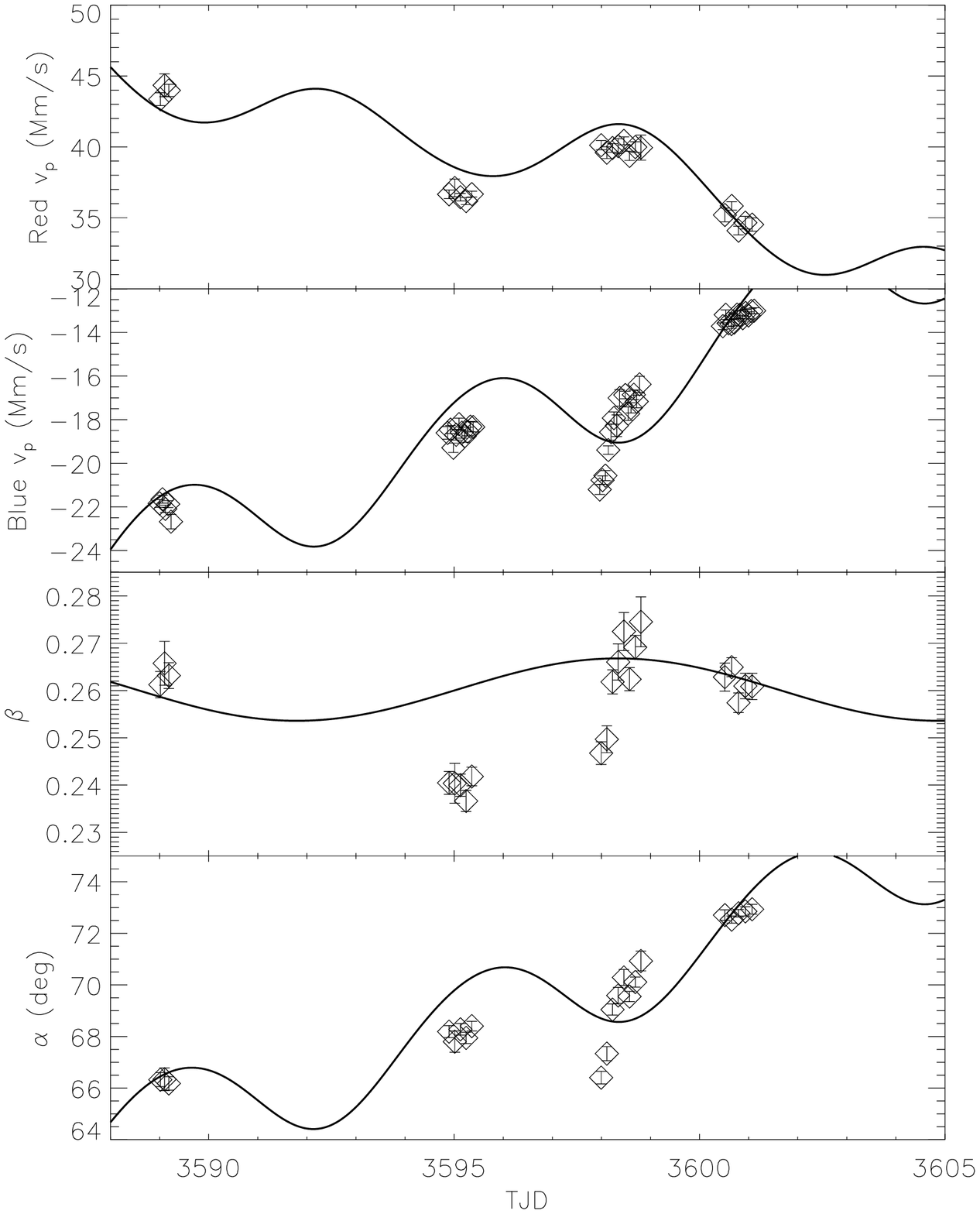}
\caption{Doppler shifts for the red- and blueshifted X-ray emission lines and derived quantities.
From the top, the panels show the Doppler shifts of the red jet
(projected along the line of sight $v_p$), the projected Doppler velocities of the blue
jet, the inferred jet speed relative to $c$, and the inferred angle to the line of sight ($\alpha$).
The smooth curves give the expected values for the nominal ephemeris for
the kinematic model, with a precession delay of 4 d, as in Fig.~\ref{fig:optxray}.
The values of $\alpha$
and $\beta$ are computed under the assumption that the blue and red
jets are directly opposed with the same speed
(see Eqs.~\ref{eq:gamma} and \ref{eq:alpha}).
The correlation of the computed jet parameters, $\alpha$ and $\beta$
on TJD 3598 appear to result from a breakdown in these assumptions, rather
than true velocity and jet angle correlations because the Doppler shift of the red
jet does not vary on that day.}
\label{fig:doppler}
\end{figure}

\noindent
Shown in Fig.~\ref{fig:doppler} are the results from calculations
using Eqs.~\ref{eq:gamma} and \ref{eq:alpha}, from which the jet
velocity and angle to the line of sight can be determined
under the assumption that the jets have the same speed
and opposite directions.
Given that only one
of the two jets' Doppler shifts shows a significant trend near TJD 3598
while both of the derived parameters increase, it seems unlikely that
both of these two assumptions are true, so we conclude that the
jets' speeds or directions change independently on time scales
of order 25 ks or less.

In order to determine which physical characteristics of the two jets differ
on short time scales, we go back to the kinematic model
and adjust its four parameters:
the jet precession phase $\phi$, precession angle $\theta$,
$\beta$, and the delay between a measurement of the X-ray Doppler shift
and that measured in the optical band.
The precession and nutation periods were fixed, along with the
nutation phase and amplitude and parameters associated with
the slight velocity variation with orbital period found by
\citet{2005ApJ...622L.129B} (observable in the panel showing $\beta$
in Fig.~\ref{fig:doppler}).
The predicted variations are obtained by differentiating the
equations for the Doppler shifts, holding all other parameters fixed.
Coincidentally, the
derivative of the jet blueshift with respect to either $\theta$
or $\beta$ is very nearly zero at the end of the observation
period (TJD 3602), so that the observed blueshift deviations
are insensitive to modest changes of these two parameters.

\begin{figure}
\centering
\includegraphics[width=12cm]{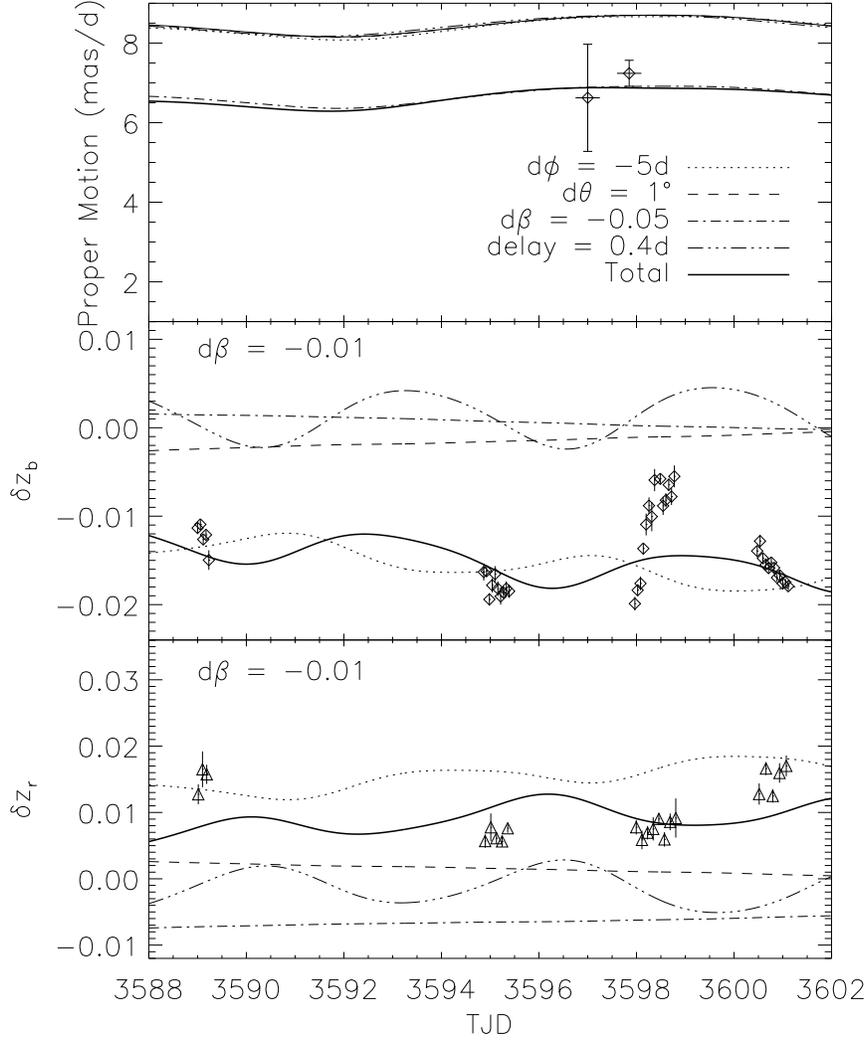}
\caption{
 Comparison of the jet proper motion and the Doppler shift deviations
 from the nominal ephemeris to models with small adjustments
 of four parameters: precession phase $\phi$, precession angle $\theta$,
 $\beta$, and an X-ray to optical time delay (see text).
 {\em Top:} Proper motion of the jet knots.  Fitting the data requires that
 $\beta$ decrease by 0.05; changes to other parameters have little effect.
 {\em Middle:} Deviation of the blueshift from the nominal ephemeris.
 The systematic offset of about 0.015 is obtained only
 by a precession phase shift.
 Changing $\beta$ or $\theta$ has little effect.  Parameter adjustments
 are the same as in the top panel except for d$\beta$.
 {\em Bottom:} Deviation of the redshift from the nominal ephemeris.
 Even after making adjustments to the kinematic model, significant
 Doppler shift residuals remain.
}
\label{fig:residuals}
\end{figure}

In Fig.~\ref{fig:residuals}, we show our best fit using small
parameter adjustments.
The position angle of the VLBA jet is very close
to the expected value: 122$\arcdeg$
at TJD 3597.2.  The expected VLBA proper motion, 8.6 mas/day, requires a
change of $\beta$ of -0.05 but changing other kinematic model
parameters barely affects the predicted proper motion.
While the blueshifts determined from the X-ray data are insensitive
to d$\beta$ values as large as 0.05, the X-ray spectra's redshifts rule
out values larger than 0.01.

Given that the knots appear to move ballistically,
the low proper motions could be explained in this model
if the jet decelerates by 20\% between
the location of the X-ray emission to where the VLBA knots appear about
1 day later, at about $5 \times 10^{14}$ cm from the core.
However, the optical emission arises at about the same distance as
the radio knots but the
optical Doppler shifts agree well with those derived from the X-ray spectra,
so this explanation seems unlikely.
Alternatively, the distance
to SS 433 could be 20\% smaller than assumed: 4.5 $\pm$ 0.2 kpc, which is
consistent with the value determined by \citet{2002MNRAS.337..657S}.

The systematic deviations of the blueshifts by about $-0.015$ can
only be accommodated in this approach by a precession phase
shift of about -5 d (out of 162.5); it is as though the blueshifts are
given by the model from 5 d prior (cf. Fig.~\ref{fig:doppler}).
The redshift deviations indicate a comparable phase shift
but the systematic offset of about $+0.010$ could be explained
by a combination of other parameter changes
(e.g., by d$\beta = -0.02$ if d$\phi = 0$). 
Even with perturbations from the kinematic model,
there are significant Doppler
shift residuals up to 0.05 from the fit.
The largest of these occurs near TJD 3598, reinforcing
the conclusion that the kinematic model's
assumptions do not hold on a time scale of 0.1 d.

\section{X-ray Spectra}
\label{sec:spectra}

The grating data were obtained from TGCat and reprocessed to reduce the
spectral extraction widths because the default MEG region (100 pixels)
overlaps the dispersed HEG spectrum above 7 keV, so that the high energy
events are not properly assigned to the HEG spectrum.
After event assignment, the source extraction region was set to the
standard width used by TGCat ($\pm$ 2.39\arcsec).
Background was selected from two regions from 2.39-7.20\arcsec\ from the dispersion.
The MEG background spectrum was generally negligible
for 1.6 \AA\ $< \lambda < 18$ \AA\ (0.69-7.75 keV), and
is dominated by mirror scattering (consisting of $< 2$\% of the total flux), so is
ignored for most analysis.
Background is negligible for the entire HEG range of interest (1.25-16 \AA\ or
0.77-10.0 keV) and is also dominated by mirror scattering at a similar level
as in the MEG spectrum.


\subsection{Eclipse}
\label{sec:eclipse}

During eclipse, part of the X-ray emission is blocked, so we developed
a spectrum of the occulted region by comparing to the non-eclipsed spectrum.
Based on the {\it RXTE} light curves (Fig.~\ref{fig:xte}), we
expect the eclipsed spectrum to
be dominated by flux above 4 keV.
We first combined the HETGS data taken outside of eclipse after
correcting them for the changing blueshift using the data shown in
Fig.~\ref{fig:trailed} where several strong lines are used to determine
the mean blueshift in 5 ks intervals.  See \S \ref{sec:dopplershifts}
for more details.  These mean blueshifts were
then fit to low order splines to obtain a smoothly varying function that
could be applied to each X-ray event before binning into a spectrum.
In order to avoid spurious features when comparing blueshifted
jet spectra, four small wavelength ranges were
excised from the data and response functions corresponding to
unredshifted Fe~{\sc I} and the strongest redshifted lines: Fe~{\sc XXV},
S~{\sc xvi}, and Si~{\sc xiv} before combining in the reference spectrum.

\begin{figure}
\center
\includegraphics[width=11cm,angle=90]{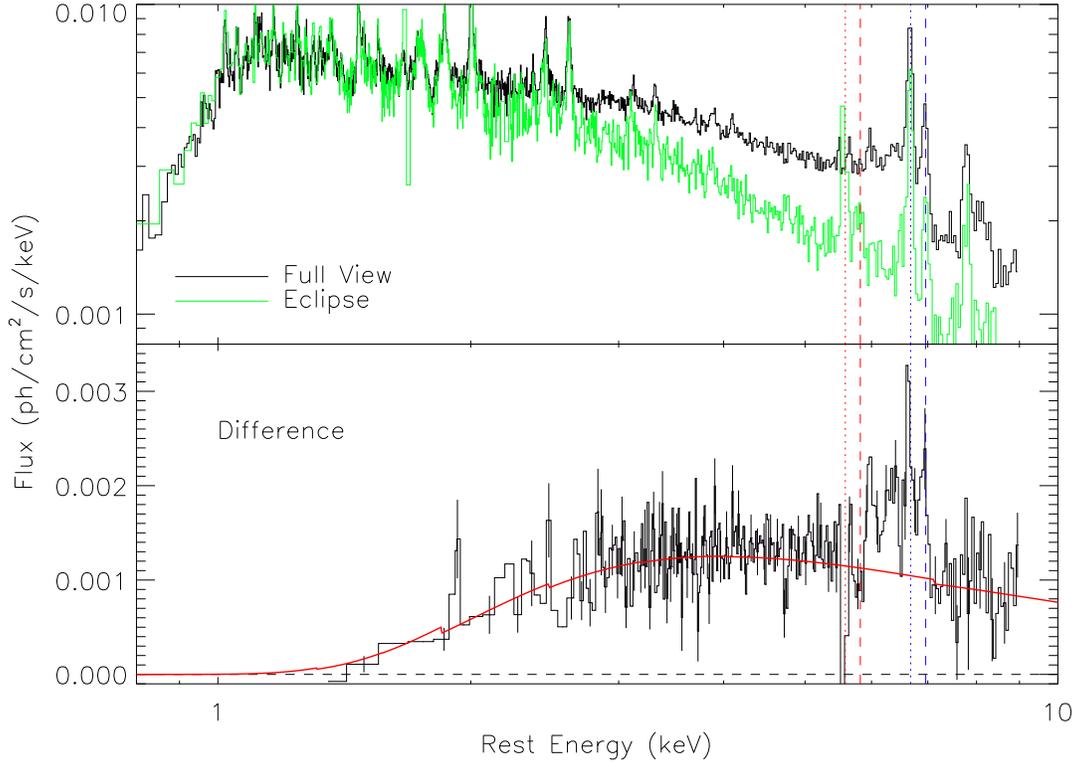}
\caption{Chandra HETGS spectra of SS 433.  {\it Top panel:} Spectra outside of eclipse (black)
and during eclipse (green).  {\it Bottom panel:} Difference between spectra in the top panel,
giving a spectrum of the eclipsed emission region; the model
(solid red line) is a simple continuum,
described in the text.  Note that the continuum below 2 keV and
lines below 3 keV in the blue-shifted jet spectrum are almost completely
cancelled in the subtraction.  The two
spectra in the top panel were corrected for the blue-shifted jet's Doppler shift over several
observations; the red-shifted jet's lines of Fe~{\sc xxv}, S~{\sc xvi}, and Si~{\sc xiv} were excised from
the full view spectra but a few weaker lines (e.g. Fe~{\sc xxvi})
from the red-shifted jet in the non-eclipse spectra
may show up in the difference spectrum.
Vertical lines mark the red-shifted (red) or blue-shifted (blue) lines
of Fe~{\sc xxv} (dotted) or Fe~{\sc xxvi} (dashed), as they are found in the eclipse
spectrum; after excising the red-shifted Fe~{\sc xxv} lines from the full view spectra
before correcting for their blue-shifts and combining,
there is a corresponding negative residual in the difference spectrum. }
\label{fig:eclipse}
\end{figure}

The source didn't vary significantly over the course
of the campaign (Fig.~\ref{fig:lc2005}) and the spectra from non-eclipsed observations
are nearly identical, so they are combined into one average non-eclipsed spectrum.
Performing the same operation on the eclipsed data, we find that
the resultant spectrum matches the non-eclipsed spectrum
almost exactly below 2 keV (Fig.~\ref{fig:eclipse}).
Importantly, the blue jet lines in the composite spectrum match
those of the eclipse spectrum to better than 10\%.
Thus, it appears that the jets haven't changed physically over the course
of the campaign, spanning about 10 days, and that the
blueshifted jet's cooler portions are fully visible during the eclipse.

As expected, the difference spectrum (Fig.~\ref{fig:eclipse}) is dominated by a hard
continuum.
Some of the features may be due to incomplete subtraction
of weak red jet lines (e.g. Fe~{\sc xxvi}) or continuum.
Also shown in Fig.~\ref{fig:eclipse} is a cutoff power law
continuum model of the occulted flux:
\begin{equation}
n_E = A (E/E_0)^{-\Gamma} e^{-N_H \sigma(E)} e^{-E/E_b}
\end{equation}
\noindent
where $A = $ 0.0016 ph cm$^{-2}$ s$^{-1}$ keV$^{-1}$, $E_0 = 6$ keV
is an arbitrary reference energy,
$\Gamma = 0.4$ is the photon spectral index,
$N_H = 3 \times 10^{22}$ cm$^{-2}$ is the column density of cold
gas along the line of sight, $\sigma(E)$ is the cross section to photoionize
neutral gas with cosmic abundances, and $E_b = 20$ keV
is the break energy.
Most of the blue Fe~{\sc xxv} and Fe~{\sc xxvi} jet lines also appear to subtract out
in this method of comparing the spectra.
At most, 20\% of the jet that provides the blue jet's Fe~{\sc xxv} line is blocked.
Up to 30\% of the blue jet's Fe~{\sc xxvi} could be blocked as well.
Thus, we confirm the result of \citet{2006lopez} that these two lines
of the blueshifted jet are mostly visible during eclipse.
It is somewhat harder to determine how much of the red jet is blocked
because it was somewhat fainter than the blue jet during this campaign,
when there was a large difference in the blue and red Doppler shifts.
However, models of the 2001 eclipse spectra indicate
that the (somewhat blended) Fe~{\sc xxv} and Fe~{\sc xxvi}
lines from both jets were similarly bright during the 2001
eclipse observations \citep[][and Fig.~\ref{fig:4spectra}]{2006lopez}.

\subsection{Modelling the Jet Emission Line Fluxes}
\label{sec:models}

\begin{figure}
\centering
\includegraphics[angle=270,width=15cm]{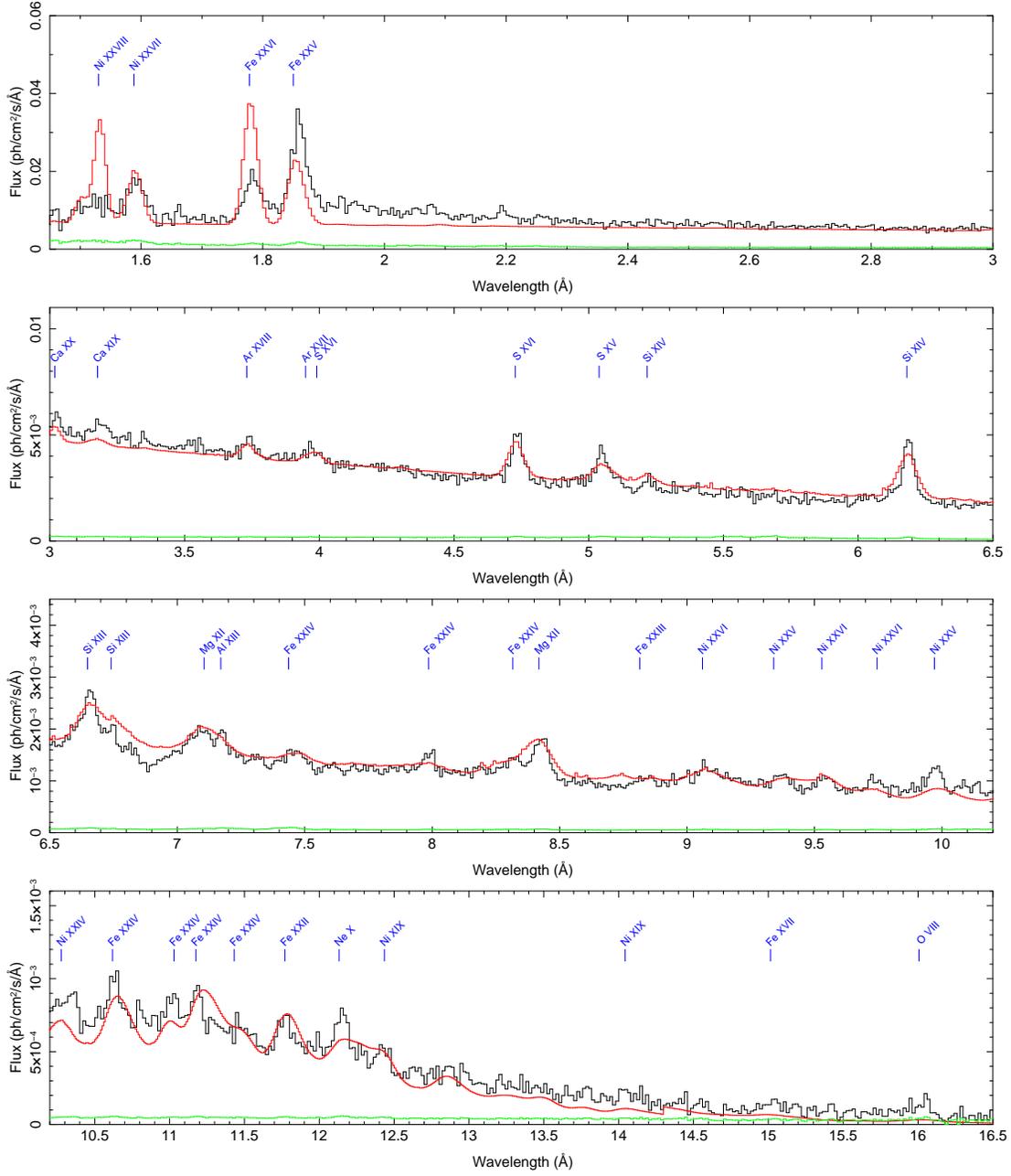}
\caption{
 X-ray spectrum formed from the HEG data (top panel only)
 or combining the HEG and MEG data
 after correcting for the Doppler shift of the blueshifted jet.
 {\em Green line:} Statistical uncertainties in the flux measurements.
 {\em Red line:} Four temperature plasma model providing
 a generally adequate fit to the spectrum.  Residuals near 2 \AA\ are
 primarily due to the redshifted jet's
 continuum, which is somewhat weaker than those of the blueshifted jet.
 Line identifications are labeled where there are features in the
 spectrum and confirmed by the model.
}
\label{fig:apedfit}
\end{figure}

As in \S\ref{sec:eclipse}, we combined all the X-ray spectra after
correcting them for the changing blueshift in order to obtain a spectrum
of the blueshifted jet.
In this case, however, we also included the data from the eclipse observation
with $E < 2.2$ keV, which is unaffected by the occultation, as found in \S\ref{sec:eclipse}.
As in Paper I, the HEG and MEG spectra were modeled jointly
using {\tt ISIS} and the APED atomic data base of line
emissivities and ionization balance.
Starting from the model from Paper I, a moderately good fit was obtained to
the line flux data with a four component model but based on a different temperature
grid ($\times 2$ intervals in $T$), and at somewhat higher temperatures than
found in Paper I.
See Fig.~\ref{fig:apedfit} for a
comparison of the model with the HETGS data.
Emission measures from the thermal model are given in Table~\ref{tab:aped}.
Other fit parameters were the turbulent velocity, $\sigma =1800$ km/s,
the metal overabundance factor, 6.2, the Ni overabundance factor, 89, and
the interstellar absorption column density, $1.165 \pm 0.006 \times 10^{22}$ cm$^{-2}$.

We find no radiative recombination continuum
features in these spectra and no other
evidence for photoionization, in contrast with Paper I.
The 9.10\AA\ line was
misidentified as the Ne {\sc x} radiative recombination continuum feature
in Paper I, which we now identify with a Li-like Ni emission line.
The misidentification resulted in part due to an early atomic data
base with insufficient Ni lines.  With these lines included and a solid detection
of He-like Ni at 1.592 \AA\ (rest frame), it is now clear that Ni lines are
readily found because Ni is highly
overabundant  -- by $\times 15$ relative to other metals.

\begin{deluxetable}{cc}
\tablecolumns{4}
\tablewidth{0pc}
\tablecaption{Blue Jet Parameters from a Multi-Temperature Model\tablenotemark{a}
\label{tab:aped} }
\tablehead{\colhead{$T$} & \colhead{$EM$}  \\
  (10$^6$ K) & (10$^{57}$ cm$^{-3}$) }
\startdata
   12 &  0.66 $\pm$  0.02  \\
  24 &  0.62 $\pm$  0.05  \\
 48 & $< 0.037$  \\
 96  & 3.3 $\pm$  0.2  \\
 192  & $< 0.054$ \\
 384  & 22.8  $\pm$  0.3   \\
\enddata
\tablenotetext{a}{Uncertainties and limits are for
the 90\% confidence level  and 6 interesting parameters.}
\end{deluxetable}

The line width parameter from this fit appears to be too large
for the long wavelength lines (see Fig.~\ref{fig:apedfit}).
The HEG and MEG spectra were then jointly fit using Gaussian profiles at the
positions of lines expected from the multi-temperature plasma modeling
and observed in the data.  The line widths were held fixed in five wavelength
groups, while fitting for the line positions and normalizations.
Results are given in Table~\ref{tab:linefluxes}.
The continuum model was set to a simple power law for each group.
Line strengths were found to be comparable to those of Paper I, with
the exceptions that the Fe {\sc xxvi} and {\sc xxv} lines are about twice
as strong in the 2005 observation.

Line width measurements are given in table~\ref{tab:linewidths}.
Systematic errors due to poor continuum modeling, unresolved blends
modeled as single Gaussians, and unmodeled weak lines are estimated
to be of order 100-150 km s$^{-1}$.
Line blending, mostly important for $\lambda > 6.5$ \AA, is also
problematic for the Fe~{\sc xxv} line; this line
dominates the average width in the 1.5-3.5 \AA\ band, which
is clearly larger than other values.
The line widths are mostly somewhat larger than found in Paper I but
comparable to the two values obtained by \citet{2003PASJ...55..281N},
who claimed that the ionized Fe lines were broader than the lines at 2 keV.
Our examination of the observation they analyzed shows a noisy spectrum
and at most very weak
lines at 2 keV, to the point that they are not clearly
detected (see Fig.~\ref{fig:4spectra}, where observation 1942 dominates
the total flux).
Given that Fe~{\sc xxv} consists of several unresolved
lines and dominates the
broad line measurement of \citet{2003PASJ...55..281N} and
that the lines may have blurred as the Doppler shift changed
during the 20 ks observation,
we are not confident that the Fe~{\sc xxv} lines are significantly
broader than other lines in the spectrum.
The smallest $\sigma$
obtained here is consistent with the average obtained in Paper I,
729 $\pm$ 34 km s$^{-1}$, so we assume that this value is appropriate
in modeling the jets for the remainder of the paper.

\begin{deluxetable}{rcrcl}
\tablecolumns{5}
\tablewidth{0pc}
\tabletypesize{\scriptsize}
\tablecaption{SS 433 Blueshifted Jet Lines\tablenotemark{a}
\label{tab:linefluxes} }
\tablehead{\colhead{$\lambda_{\rm rest}$} & \colhead{$\lambda_{\rm obs}$} & \colhead{$\Delta z$}
    & \colhead{Flux} & \colhead{Identification} \\
  \colhead{(\AA)} & \colhead{(\AA)} & \colhead{}
    & \colhead{($10^{-6}$ ph cm$^{-2}$ s$^{-1}$)} & \colhead{} }
\startdata
 1.532 &  1.526 $\pm$ 0.005 &  -0.0040 $\pm$ 0.0034 & 128. $\pm$ 30. &               Ni {\sc xxviii} \\
 1.592 &  1.592 $\pm$ 0.002 &   0.0003 $\pm$ 0.0011 & 279. $\pm$ 31. &                Ni {\sc xxvii} \\
 1.780 &  1.786 $\pm$ 0.001 &   0.0032 $\pm$ 0.0007 & 272. $\pm$ 21. &                 Fe {\sc xxvi} \\
 1.855 &  1.860 $\pm$ 0.001 &   0.0030 $\pm$ 0.0003 & 713. $\pm$ 23. &                  Fe {\sc xxv} \\
 3.020 &  3.025 $\pm$ 0.005 &   0.0016 $\pm$ 0.0017 &  31. $\pm$  6. &                   Ca {\sc xx} \\
 3.187 &  3.191 $\pm$ 0.004 &   0.0013 $\pm$ 0.0014 &  48. $\pm$  6. &                  Ca {\sc xix} \\
 3.733 &  3.740 $\pm$ 0.003 &   0.0020 $\pm$ 0.0008 &  31. $\pm$  5. &                Ar {\sc xviii} \\
 3.962 &  3.970 $\pm$ 0.004 &   0.0021 $\pm$ 0.0011 &  29. $\pm$  5. &    Ar {\sc xvii}, S {\sc xvi} \\
 4.729 &  4.735 $\pm$ 0.001 &   0.0013 $\pm$ 0.0003 & 111. $\pm$  6. &                   S {\sc xvi} \\
 5.055 &  5.050 $\pm$ 0.002 &  -0.0011 $\pm$ 0.0004 &  89. $\pm$  6. &                    S {\sc xv} \\
 5.217 &  5.224 $\pm$ 0.006 &   0.0013 $\pm$ 0.0012 &  20. $\pm$  5. &                  Si {\sc xiv} \\
 6.182 &  6.186 $\pm$ 0.001 &   0.0006 $\pm$ 0.0001 & 170. $\pm$  5. &                  Si {\sc xiv} \\
 6.648 &  6.652 $\pm$ 0.002 &   0.0005 $\pm$ 0.0002 &  82. $\pm$  3. &                Si {\sc xiii}r \\
 6.740 &  6.735 $\pm$ 0.004 &  -0.0007 $\pm$ 0.0005 &  30. $\pm$  3. &                Si {\sc xiii}f \\
 7.101 &  7.083 $\pm$ 0.003 &  -0.0024 $\pm$ 0.0004 &  43. $\pm$  3. &   Mg {\sc xii}, Ni {\sc xxvi} \\
 7.173 &  7.172 $\pm$ 0.004 &  -0.0001 $\pm$ 0.0005 &  35. $\pm$  3. &                 Al {\sc xiii} \\
 7.457 &  7.459 $\pm$ 0.006 &   0.0003 $\pm$ 0.0008 &  20. $\pm$  3. &     Fe {\sc xxiii}/{\sc xxiv} \\
 7.989 &  7.984 $\pm$ 0.004 &  -0.0006 $\pm$ 0.0005 &  27. $\pm$  2. &                 Fe {\sc xxiv} \\
 8.316 &  8.309 $\pm$ 0.004 &  -0.0009 $\pm$ 0.0004 &  33. $\pm$  2. &                 Fe {\sc xxiv} \\
 8.421 &  8.425 $\pm$ 0.002 &   0.0005 $\pm$ 0.0002 &  66. $\pm$  2. &                  Mg {\sc xii} \\
 8.815 &  8.851 $\pm$ 0.010 &   0.0041 $\pm$ 0.0011 &  12. $\pm$  2. &                Fe {\sc xxiii} \\
 9.075 &  9.080 $\pm$ 0.004 &   0.0005 $\pm$ 0.0004 &  22. $\pm$  2. &                 Ni {\sc xxvi} \\
 9.372 &  9.371 $\pm$ 0.005 &  -0.0001 $\pm$ 0.0005 &  13. $\pm$  2. &       Ni {\sc xxvi}/{\sc xxv} \\
 9.529 &  9.538 $\pm$ 0.006 &   0.0010 $\pm$ 0.0006 &   8. $\pm$  2. &                Ni {\sc xxiii} \\
 9.745 &  9.737 $\pm$ 0.007 &  -0.0008 $\pm$ 0.0007 &   6. $\pm$  2. &                 Ni {\sc xxvi} \\
 9.970 &  9.973 $\pm$ 0.003 &   0.0003 $\pm$ 0.0003 &  26. $\pm$  2. &                 Ni {\sc xxvi} \\
10.634 & 10.633 $\pm$ 0.005 &  -0.0000 $\pm$ 0.0004 &  23. $\pm$  2. &                 Fe {\sc xxiv} \\
11.026 & 11.025 $\pm$ 0.007 &  -0.0001 $\pm$ 0.0006 &  18. $\pm$  2. &     Fe {\sc xxiv}/{\sc xxiii} \\
11.176 & 11.182 $\pm$ 0.004 &   0.0005 $\pm$ 0.0004 &  24. $\pm$  3. &                 Fe {\sc xxiv} \\
11.432 & 11.465 $\pm$ 0.017 &   0.0029 $\pm$ 0.0015 &   6. $\pm$  2. &                 Fe {\sc xxiv} \\
11.753 & 11.778 $\pm$ 0.008 &   0.0021 $\pm$ 0.0007 &  16. $\pm$  3. &     Fe {\sc xxiii}/{\sc xxii} \\
12.134 & 12.146 $\pm$ 0.006 &   0.0010 $\pm$ 0.0005 &  35. $\pm$  4. &                    Ne {\sc x} \\
12.435 & 12.426 $\pm$ 0.010 &  -0.0007 $\pm$ 0.0008 &  16. $\pm$  3. &                  Ni {\sc xix} \\
14.060 & 14.109 $\pm$ 0.024 &   0.0035 $\pm$ 0.0017 &   9. $\pm$  3. &                  Ni {\sc xix} \\
15.014 & 15.021 $\pm$ 0.029 &   0.0004 $\pm$ 0.0020 &   5. $\pm$  2. &                 Fe {\sc viii} \\
16.006 & 16.022 $\pm$ 0.013 &   0.0010 $\pm$ 0.0008 &  14. $\pm$  3. &                  O {\sc viii} \\
\enddata
\tablenotetext{a}{Rest wavelengths are computed for blends by applying weights
  equal to the fractional flux contribution to the blend, according to the multi-temperature
  plasma model.}
\end{deluxetable}

\begin{deluxetable}{ll}
\tablecolumns{2}
\tablewidth{0pc}
\tablecaption{Blueshifted Jet Line Widths
    \label{tab:linewidths} }
\tablehead{\colhead{Wavelength Range} & \colhead{$v$\tablenotemark{a}} \\
   \colhead{(\AA)} & {(km s$^{-1}$)} }
\startdata
1.5 -- 3.5	&	1900 $\pm$ 100 \\
3.5 -- 6.5	&	1100 $\pm$  40 \\
6.5 -- 9.0	&	1260 $\pm$  40 \\
9.0 -- 11.5	&	760 $\pm$  40 \\
11.5 -- 16.0 &	950 $\pm$  80 \\
\enddata
\tablenotetext{a}{Velocity width ($\sigma$) of Gaussian line profiles.
Uncertainties are statistical only; systematic errors of 100-150 km s$^{-1}$
may be present due to line blends, continuum modeling, and unmodeled
weak lines.}
\end{deluxetable}

\begin{figure}
\centering
\includegraphics[width=14cm]{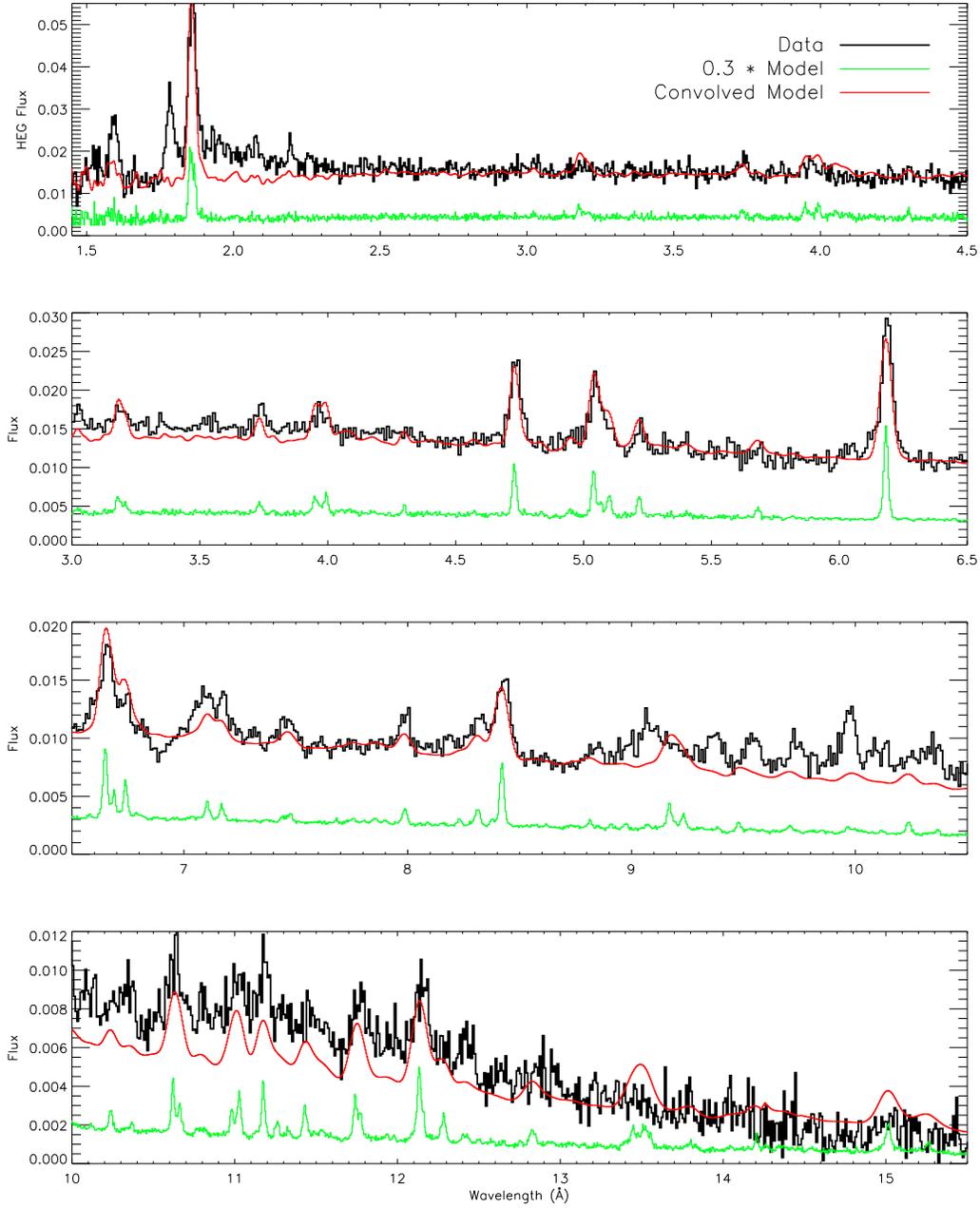}
\caption{
 X-ray spectra of SS 433
 after correcting for the Doppler shift of the blueshifted jet,
 compared to a model based on $\theta^1$ Ori C.
 {\em Green line:} Original spectrum of $\theta^1$ Ori C,
 modified as described in the text and then scaled
 down by a factor of 0.3.
 {\em Red line:} Spectrum of $\theta^1$ Ori C, as modified
 and then Doppler broadened by a Gaussian with $\sigma = 1000$ km/s.
 The match to the SS 433 spectrum is generally good, with notable exceptions:
 SS 433 shows a much stronger Fe {\sc xxvi} line at high temperature and
 Ni is highly overabundant (see Fig.~\ref{fig:apedfit}).
 Also the low temperature lines of Mg {\sc xi} (9.2 \AA) and Ne {\sc ix} (13.5 \AA)
 are not observed in SS 433.
}
\label{fig:orion}
\end{figure}

A close proxy to the spectrum of SS 433 was identified in the line-rich
spectrum of $\theta^1$~Ori~C \citep{2003ApJ...595..365S}, the brightest
young O star in the Orion Nebula's Trapezium.
A spectrum was derived from 12 observations in
the TGCat archive (obsIDs 3, 4, 2567, 2568, 7407-12, 8568, and 8569),
totaling 398 ks of exposure, and then simply modified to
provide a close match to the SS 433 continuum and overall line strengths
by $f_{\rm model} = C_1 f_{\theta^1 \rm{Ori C}} E^{0.5} + C_2$ and then
applying the ISM absorption correction, where $E$ is energy in keV, and $C_1$
and $C_2$ were constants (6.5 and 0.0085 ph/cm$^2$/s, respectively).
The modified spectrum was
then Doppler broadened by a Gaussian with $\sigma = 1000$ km/s.
The match to the SS 433 spectrum is generally good, as seen
in Fig.~\ref{fig:orion}.
There are notable exceptions: SS 433 shows
the high temperature lines of Fe {\sc xxvi} (1.78 \AA) and Ni {\sc xxvii} (1.60 \AA).
Ni is highly overabundant, as found in the APED fit and evidenced by the
Ni {\sc xxvi} and Ni {\sc xxv} L lines in the 9.1-10.0 \AA\ range
(see Fig.~\ref{fig:apedfit}).
Finally, the low temperature lines of Mg {\sc xi} (9.2 \AA) and
Ne {\sc ix} (13.5 \AA) seen in $\theta^1$ Ori C are not observed in SS 433.
Thus, the spectrum of  $\theta^1$ Ori C is quite similar to that of SS 433
except that SS 433 has more high temperature gas, less gas at low temperatures,
highly overabundant Ni relative to Fe,
and more substantial Doppler broadening.

The comparison to $\theta^1$~Ori~C generally indicates that
the X-ray emission is due to thermalized gas in both the O star and
in the SS 433 jets.
\citet{2003ApJ...595..365S} and \citet{2005ApJ...628..986G} provide
a physical picture of the O star's
spectrum using a magnetically channeled wind shock, based on
papers by \citet{1997ApJ...485L..29B,1997A&A...323..121B}
and \citet{2002ApJ...576..413U}.
This model provides high temperatures for the emission measure
distribution and some Doppler broadening by turbulence is expected.
We suggest that shocks provide the basic heating mechanism in
SS 433 as well.  It seems reasonable that a higher temperature
component is required for SS 433 than is found in the O star wind,
due to the mildly relativistic speeds of the jets. 
Of course, in SS 433, the heated gas is moving outward in opposed
jets at 0.26$c$ relative to
the compact object in the system, so any shock-heating would take
place in the frame of material already moving at high speed.
The similarity of the SS 433 spectrum to that of an O star does
not provide an indication of the type of companion star, which
is still a matter of some debate.

\section{Discussion and Summary} 
\label{sec:discussion}

\subsection{The Rapid Doppler Shift Change}

The Doppler shifts of the blue jet lines are observed to undergo a rapid change
during the 2005 observations with the {\em Chandra} HETGS.
The change
of $\sim 5000$ km s$^{-1}$ or $\Delta z = 0.017$
occurs on a time scale of 25 ks, much shorter than the
time scale than any of the known periodicities
in the system due to precession, orbit, or nutation.
Because the blueshifted jet's Doppler variations are
insensitive to jet speed variations during the observation,
we infer that the jet's direction was affected locally.
Moreover, the emission lines are not appreciably broadened
temporally -- if the cooling time of the jet were much longer than
5 ks, then the variation observed in Fig.~\ref{fig:trailed} would be smeared
horizontally, which is not observed.  Thus, most of the gas
must have cooled in less than 5000 s.

No large Doppler shift change was observed at the same
time in the redshifted jet, indicating that the change is due to a local effect near the
point where the heated plasma is accelerated to 0.26$c$.
One interpretation
of these data is that the jets' terminal velocities are determined
by environmental effects that perturb the direction of the jet.
Velocity perturbations alone are insufficient to explain the large Doppler shift
changes -- either the -0.015 systematic offset from the kinematic model or
the 0.05 residuals that remain after adjusting the parameters of the kinematic
model.  Angular deviations of order 3$\arcdeg$ along the direction
of precession would be sufficient to explain the residuals of 0.05
in the blueshifted jet's Doppler shift.

From the concurrent VLBA monitoring, a jet knot was found that
was ejected 0.22 $\pm$ 0.05 d before the start of observation 5514
and it appears that the blueshifted jet had already changed direction
as the observation started.
With a possible systematic uncertainty of about 0.25 d in the
ejection time, it is possible
that these two events are causally related.
Radio emitting knots were observed for several days, maintaining
their discrete nature.  By contrast,
the X-ray emitting region is continuously visible,
smoothly shifting, and dominated by a single Doppler component.
Thus, it seems that the SS 433 jets go through a transition
from continuous flow to regularly spaced clumps
-- X-rays are emitted during the continuous flow, while the discrete knots
produce most of the optical and radio emission.  Such
clumping could occur as a result of a thermal cooling instability
\citep{1988A&A...196..313B} or by slight
variations in jet speed where slow moving
flow is caught by faster moving material, perhaps causing shocks
that reheat the gas and accelerating particles that produce radio
emission via synchrotron radiation \citep[e.g.][]{2005MNRAS.358..860M}.
Unfortunately, there was no HETGS
coverage during the first knot ejection and the chances are fair
that the 75 ks exposure of observation 5514 elapsed before
the ejection of another radio knot.

\subsection{Locating the Jet Origin}

The location of the x-ray emission within the binary system can
be constrained using various measurements from these observations.
First off, limiting the cooling time to less than 5000 s requires that
the X-ray emitting gas moving at 0.26$c$ be less
than $4 \times 10^{13}$ cm from
the point at which the jet direction is set.

The similarity of the X-ray and optical Doppler shifts limits the
time (and distance) between the X-ray emission and the optical
emission to less than about 0.4d (\S \ref{sec:dopplershifts}), or 
$< 2.7 \times 10^{14}$ cm.
Thus, it appears that there
is no significant deceleration of the jet between the
optical and X-ray line emission regions.
Combining with the distance limit from the jet base to the
X-ray emission, the optical emission should occur
$< 3.1 \times 10^{14}$ cm from the jet base.
Note that  \citet{1999A&A...351..156P} obtained
a similar estimate of the distance to the optical emission region:
$4 \times 10^{14}$ cm.

We limit the location of the hard X-ray emission using the eclipse observation.
The source spectrum was remarkably steady during the 2005 campaign,
so that it is straightforward to compare the spectrum during eclipse to other periods
during the orbit.
It is clear that the cool portions of the X-ray emitting jet are not occulted during the
eclipse.
The difference spectrum (Fig.~\ref{fig:eclipse}) shows that the blocked emission
is harder than the average, unblocked emission.
The RXTE data \citep[\S~\ref{sec:xte} and ][]{2011ApJS..196....6L} bolster this result:
the eclipse depth increases with energy to the point that over half the flux
above 15 keV is occulted during eclipse while less than 20\% is blocked
in the 2-4 keV band.
Thus, the star is larger than
previously estimated by \cite{2006lopez} and large enough to block the
highest temperature gas of both the red- and the blueshifted jets of the
hardest emission during mid-eclipse.
The eclipse is not detected when the jets are nearly perpendicular
to the line of sight and the average X-ray count rates are similar
to that in eclipse \citep{2002ApJ...566.1069G}, suggesting self-obscuration
by a flared disk at such precessional phases.  The outer edge of
the compact object's accretion disk could be optically thick to hard X-rays
and very thick spatially, so that its angular size is comparable to the
size of the companion star, as subtended from
the compact object.\footnote{\citep{2002ApJ...566.1069G} suggested
that a circumbinary disk is responsible for obscuring the central X-ray source
when the accretion disk is viewed edge-on but in this case the X-ray flux
should be {\em lower} than during eclipse, as a larger region is obscured.}
\citet{2011ApJS..196....6L} found mid-eclipse
to be at TJD 2745.69 $\pm$ 0.18, which corresponds to an orbital
phase of $0.080 \pm 0.021$ -- delayed with respect to the optical
light minimum.  The effect was first noticed
by \citet{1989PASJ...41..491K} and a similar
result was found by \citet{2002ApJ...566.1069G},
who suggested that the companion has an extended cometary
tail of optically thick gas trailing behind it in the orbital plane.
Taken with the observation that the redshifted
Fe {\sc xxv} line has nearly identical flux as that of the blueshifted
jet during eclipse and near the precessional ``cross-over'' point
when the jets are nearly perpendicular to the line of sight
\citep[][and Fig.~\ref{fig:4spectra}]{2006lopez},
we conclude that this emission must occur farther than
the donor star's Roche radius,
$\sim 2 \times 10^{12}$ cm from the disk plane,
for binary mass values reported by \citet{2008ApJ...676L..37H}.
This distance is comfortably smaller than that derived from
the cooling time limit of $4 \times 10^{13}$ cm.

We now place bounds on the electron density at the base of the jet.
In order for the hottest line-producing component of the plasma
to be in thermal equilibrium and visible during eclipse,
the recombination time should be shorter and the
radiative cooling time longer than the jet flow time to $10^{12}$ cm.
Based on the abundances from the plasma
fits, the cooling time for gas at $4 \times 10^8$K is
$1000/n_{12}$ s, where $n_{12}$ is the electron
density in units of $10^{12}$ cm$^{-3}$.
From the emission measure of this component, only when
$n_{12} < 10$ is the cooling time
long enough that the Fe {\sc xxvi} is still visible during eclipse.
For higher densities, the point at which the gas is heated to X-ray
temperatures must be off of the disk in order to be visible during eclipse.
The recombination time for Fe {\sc xxvi} is $\sim 2/n_{12}$ s,
so requiring that this time be short compared to the flow time
gives $n_{12} > .01$.  Thus, the electron density
should be in the range $10^{10-13}$ cm$^{-3}$ to satisfy both
requirements.

\subsection{UV De-excitation of the Si {\sc xiii} Triplet}

In Paper I, we fit the profile of the Si {\sc xiii} triplet in order to determine
the resonance, intercombination, and forbidden line fluxes.
The low value of the forbidden/intercombination line ratio was
then used to estimate the density of the
jet gas where Si {\sc xiii} is prevalent, obtaining a value of
$n_e \sim 10^{14}$ cm$^{-3}$.  This value is substantially higher
than estimated in the previous section.  We now examine the
possibility that this ratio is not
a reliable density estimate for SS 433 because excitation by
UV disk photons
can suppress the forbidden line by depopulating the upper state
before decay to the ground state.  This
mechanism has been invoked to explain the triplet line strengths
of O stars \citep[e.g.][]{2001A&A...365L.312K} and we follow a similar approach.
Atomic data were taken from the on-line NIST Atomic Spectra
Database \cite{nist.asd}.

The transition decay rate for the Si {\sc xiii} forbidden line (from 2 $^3$S
to 1 $^1$S) is $A_f = 3.56 \times 10^5$ s$^{-1}$ \citep{1977PhRvA..15..154L}.
UV excitation depends on the photon flux, $F_{\nu}$, from the accretion disk
at any of three transitions from 2 $^3$S$_1$ to 2 $^3$P$_J$, where $J$ is 0, 1, or 2.
The transitions have
wavelengths $\lambda_J = $ (878.58, 865.14, 814.71) \AA,
upper level statistical weights $g_{2,J} = (1,3,5)$, and Einstein coefficients
$ A_{21,J} = (1.41, 1.391, 1.386) \times 10^8$ s$^{-1}$ \citep{kelleher:1285}.
The excitation rate is 

\begin{equation}
\label{eq:re}
R_e =  \int \frac{F_\nu}{h \nu} \sum_J \sigma_{\nu,J} d\nu
  \approx  \sum_J \frac{\lambda_J F_\nu(\lambda_J)}{h c} \int  \sigma_{\nu,J} d\nu
  = \frac{1}{8 \pi h c^2 g_1} \sum_J g_{2,J} A_{21,J} F_J \lambda_J^5
\end{equation}

\noindent
where $g_1 = 3$ is the statistical weight of the lower level,
$\sigma_{\nu}$ is the transition's absorption cross section,
and $F_J$ is the flux density (in erg cm$^{-2}$ s$^{-1}$ \AA$^{-1}$) of the disk
at $\lambda_J$.

\citet{2002ApJ...566.1069G} provided a range of disk
temperatures for different values of the optical extinction of $A_V$, fitting
blackbody spectra to FUV measurements from HST, UV values
from \cite{1997A&A...327..648D}, and optical data
from \cite{1986ApJ...308..152W}.
They suggest that a model with $T = 21000$ K and $A_V = 7.8$
is consistent with the observed UV and optical fluxes.  However, they
also say that a model with $T = 49000$ K and $A_V = 8.2$
is marginally consistent with the data while
a model with $T = 72000$ K and $A_V = 8.4$
would be inconsistent with the data.  For illustrative purposes,
we assume a set of values: $A_V = (7.8, 7.8, 8.2, 8.4)$ for
$T_4 \equiv T/(10^4 {\rm K}) = (2.1, 4.5, 4.9, 7.2)$.
Substituting the blackbody model into Eq.~\ref{eq:re}, then

\begin{equation}
R_e = \frac{1}{4 g_1} \sum_J \frac{g_J A_{21,J}}{e^{hc/(k \lambda_J T)} - 1}
\end{equation}

\noindent
at the surface of the disk.  For the four temperatures, we find
$R_e / A_f = (0.1, 1.2, 11, 37)$, indicating that the UV flux is insufficient
to depopulate the upper state of the forbidden line transition at the
disk surface if its
temperature is below $4.5 \times 10^4$ K but is large enough for
higher disk temperatures -- i.e., level depopulation by the FUV
continuum is possible if $T/(10^4 K)$ is in the range  4.5-4.9, consistent
with the FUV-optical spectra.
As found in the last section, the jet's X-ray emission should be about
$2 \times 10^{12}$ cm from the disk surface; we find (projected) disk radii
in the range of $2-3 \times 10^{12}$ cm, based on the dereddened optical fluxes
\citep[using dust correction by][and $E(B-V) = A_V / 3.1$]{1999PASP..111...63F},
so dilution should not be significant.
Given the uncertainty in the true disk temperature, we conclude
that UV pumping may well be responsible for the low
forbidden/intercombination  line ratio found in Paper I, which led
to an electron density estimate that was too high, $\sim 10^{14}$ cm$^{-3}$
for gas with a temperature of order $10^7$ K.

\subsection{Ni Overabundance}

The APED spectral fit and the comparison to the O star spectrum both
indicate that Ni is highly overabundant.
The APED fit (and our results from Paper I) indicate that all metals
(particularly Fe) are overabundant by about a factor of 6
but the Ni abundance is $\times 15$ larger.
We consider two possibilities for the source of Ni.

One possibility is that the excess Ni is inherent in the accreted
gas, which is entrained in the jet via an interaction with the disk or
companion's wind.
This scenario requires that Ni is overabundant in the atmosphere of
the companion.
While \citet{2008ApJ...676L..37H} did not find Ni lines in the optical
spectra they obtained of the companion star,
no other elements appear to be particularly
overabundant, including Fe.  On the other hand, the
companion spectra are diluted by emission from the disk and this
dilution factor is not known {\it a priori} but determined by scaling
the line strengths.
The companion could be a
post-AGB (asymptotic giant branch) A star like BD $+48$\arcdeg\ 1220,
which was found to have a
super-solar abundance of Ni \citep{2007ARep...51..642K}.

It also seems plausible that the excess Ni originated in the
supernova (SN) that created the compact object.
The SN would have
deposited enriched material on the surface of the companion
(perhaps as fall-back after the initial blast wave).
However, in many SN models, most Ni
would have been in the form of radioactive $^{56}$Ni, which has
a half-life of 6.1 d -- much too short to be detected today, given that
the system has been observed for at least 50 yr.
For this model to apply, the SN would have to be of a rare type
that produces as much
stable Ni as Fe.  Interestingly, Ni does seem to be $\ga 10\times$
overabundant in the Crab nebula
\citep{1983A&A...127...42D,1984ApJ...281..644H,1986PASP...98.1044H,1990ApJ...354L..57H}.
\citet{2009ApJ...695..208W} developed such a SN model, starting
with an 8.8 M$_\sun$ AGB star and invoking significant electron capture, that
generates a Ni/Fe ratio that is 20 times the solar value due to the
weak production of $^{56}$Ni.

Another possibility is that Ni is formed {\it in situ} via an
interaction of a jet with the disk wind.  Such an interaction
was proposed by \cite{2006MNRAS.370..399B} as a way to
provide periodically redirected outflows from an inherently
stably pointing jet.  This scenario requires a high density
of target material, which is feasible for a wind from a supercritically
accreting disk.
One problem with this model is that there is no particular reason
to expect that only isotopes of Ni would result.
We do not detect Mn and the Ar and Ca abundances appear to
be similar to that of Fe.

\section{Summary}

The 2005 campaign to observe SS 433 simultaneously at
X-ray, optical, and radio wavelengths provided many ways to
clarify its jet production mechanism:

\begin{enumerate}
\item{A rapid change of one jet's Doppler shift was observed,
changing by 5000 km s$^{-1}$ in 25 ks, much faster than the
precessional, orbital, or nutational time scales of the system.
This rapid change may have coincided with the ejection
of a radio emitting knot, propagating ballistically away from the core.}
\item{The red- and blue-shifted jets did not
change at the same time, showing that the directions of
the two jets are independently determined or
affected by the environment.}
\item{Because the Doppler shift varied so quickly in one jet, the cooling time
of the X-ray emitting plasma was limited to $<$ 5000 s, limiting
the distance from the jet launch site to the X-ray emission region to
$< 4 \times 10^{13}$ cm.}
\item{Observing discrete features in the radio band but not in the
the X-ray spectrum shows that the flow starts out continuous but then
clumps to form knots.}
\item{Agreement between X-ray and optical Doppler shifts gives a
consistent kinematic model, in conflict with the radio blob proper motions
unless the distance to SS 433 is 4.5 $\pm$ 0.2 kpc.}
\item{The X-ray spectra were remarkably consistent before and
after eclipse, providing an opportunity to determine the X-ray
spectrum of the eclipsed emission, which is hard and mostly line-free.
Observing emission lines during eclipse requires that the jet be
longer than the Roche radius of the companion, $\sim 2 \times 10^{12}$ cm.}
\item{Modeling the jet's X-ray spectrum shows that Ni is $\times$15
overabundant relative to other metals.  We speculate that the supernova
that created the compact object in SS 433 was somewhat unusual,
having generated as much stable Ni as Fe.}
\item{Disk UV flux could affect the Si {\sc xiii} lines, possibly
invalidating the use of their fluxes as a density diagnostic in this system.}
\item{Using various arguments relating to the jet plasma model, we
derive a range for the electron density of $10^{10-13}$ cm$^{-3}$
at the jet base.}
\item{We infer that shocks excite the X-ray emitting gas to the
observed temperatures, similar to situations found in young O stars.}
\end{enumerate}

\acknowledgements

We thank the referee for excellent comments and suggestions that
resulted in an improved paper.
We thank Lydia Oskinova for pointing out the high Ni abundance in
the supergiant BD $+48$\arcdeg\ 1220.
Support for this work was provided by the National Aeronautics and
Space Administration through the Smithsonian Astrophysical Observatory
contract SV3-73016 to MIT for Support of the Chandra X-Ray Center,
which is operated by the Smithsonian Astrophysical Observatory for and
on behalf of the National Aeronautics Space Administration under contract
NAS8-03060.
The National Radio Astronomy Observatory is a facility of the National
Science Foundation operated under cooperative agreement by Associated
Universities, Inc.


\bibliographystyle{apj}                       
\bibliography{ss433}

\end{document}